\documentclass[aps,prc,twocolumn,showpacs,preprintnumbers,superscriptaddress,
                          nofootinbib,float,longbibliography]{revtex4-1}
\usepackage{color}
\usepackage{amsmath}
\usepackage[dvipsnames]{xcolor}
\usepackage{graphicx}
\usepackage{dcolumn}
\usepackage{bm}
\usepackage{appendix}
\usepackage[utf8]{inputenc}
\usepackage{multirow}
\usepackage[colorlinks=true, pdfstartview=FitV, linkcolor=red,
citecolor=blue, urlcolor=blue]{hyperref}

\usepackage{xcolor}

\def\e{{\epsilon} }

\def\P{{\mathcal P}}

\def\S{{\mathcal S}}
\def\L{{\mathcal L}}

\def\x{\bf x}
\def\p{\bf p}

\def\Eq#1{Eq.~(\ref{#1})}

\def\Fig#1{Fig.~\ref{#1}}
\def\Sect#1{Section~\ref{#1}}

\def\Tab#1{Table~\ref{#1}}

\def\be{\begin{equation}}
\def\ee{\end{equation}}
\def\bea{\begin{eqnarray}}
\def\eea{\end{eqnarray}}

\begin{document}


\title{The emission of electromagnetic radiation from the early stages of relativistic heavy-ion collisions}

\author{Jessica Churchill}
 \affiliation{Department of Physics, McGill University, 3600 University Street, Montreal, QC, Canada H3A 2T8}
\author{Li Yan}%
\affiliation{Key Laboratory of Nuclear Physics and Ion-Beam Application (MOE) \& Institute of Modern Physics\\
Fudan University, 220 Handan Road, 200433, Yangpu District, Shanghai, China}
\author{Sangyong Jeon}
\affiliation{Department of Physics, McGill University, 3600 University Street, Montreal, QC, Canada H3A 2T8}
\author{Charles Gale}
\affiliation{Department of Physics, McGill University, 3600 University Street, Montreal, QC, Canada H3A 2T8}
%




\date{\today}

\begin{abstract}
We estimate the production of electromagnetic radiation (real and virtual photons) from the early, pre-equilibrium, stage of relativistic heavy-ion collisions. The parton dynamics are obtained as a solution of the Boltzmann equation in the Fokker-Planck diffusion limit. The photon and dilepton rates are integrated and the obtained yields are compared with those from  standard sources and with available experimental data. Non-equilibrium photon spectra are predicted for Pb+Pb at $\sqrt{s_{\rm NN}} = 5.02$ TeV. 
\end{abstract}

\maketitle


\section{\label{sec:intro}Introduction}

%
The accepted theory of the nuclear strong interaction is Quantum Chromodynamics (QCD), a local gauge theory which admits a spontaneously broken chiral symmetry. This theory has been very successful in describing static properties of strongly interacting systems, and is also able to interpret and predict the outcome of high-energy scattering experiments involving hadronic particles. In spite of all its remarkable successes, there remains much to be learned about QCD. For example, the behaviour of many-body QCD in temperature and density regions far removed from equilibrium is currently  the topic of a vibrant research program. The theoretical nature of the transition between degrees of freedom belonging respectively to the partonic and confined phases has only been recently identified as a rapid crossover, occurring at $T_c \approx 150$ MeV, for zero baryon density~\cite{Ratti:2018ksb}). This region is accessible to heavy-ion experiments performed at both the Relativistic Heavy-Ion Collider (RHIC) and the Large Hadron Collider (LHC). Experiments performed at these facilities have revealed an exotic form of matter: the quark-gluon plasma (QGP) \cite{Jacak:2012dx}. 

One of the tantalizing properties of the QGP, is that - contrary to early theoretical expectations - it possesses fluid-like characteristics, and therefore can be modelled using relativistic fluid dynamics \cite{Gale:2013da}. This approach has achieved great empirical success, mainly characterized by a quantitative  interpretation of the hadronic flow systematics measured in experiments \cite{Heinz:2013th}.
Modern relativistic hydrodynamics even  enables the extraction of the transport parameters of QCD. For instance, 3D approaches now exist \cite{Schenke:2010nt}, and can be used to extract the shear \cite{Schenke:2010rr} and bulk \cite{Ryu:2015vwa,Ryu:2017qzn} viscosities of  QCD matter. 
Notwithstanding the recent progress in relativistic fluid dynamics, it is fair to write that relativistic heavy-ion collisions can still not be modelled {\it ab initio}: hybrid approaches need to be constructed. Those typically consist of an initial state followed by the hydrodynamics phase which ends with a hadronic cascade and kinetic freeze-out, when inter-particle distances exceed mean-free-paths while the interaction volume expands and cools. The cascade stage usually  relies on Monte Carlo packages where hadronic species are allowed to collide and interact with each other. A recent example is  {\sc smash} \cite{[{See, for example, }][{, and references therein.}]Weil:2016zrk}. 

The quantum nature of the initial state, especially how it evolves towards ``hydrodynamization'', is the subject of much current research. The short hydro formation times required by modern phenomenological analyses  ($\tau_0 \lesssim  1$ fm/c) is a challenge to perturbative approaches and calculations \cite{Fukushima:2016xgg}. However, initial conditions computed within the Color Glass Condensate (CGC) framework obtained using the impact parameter dependent saturation model with the classical Yang-Mills evolution of the classical gluon fields \cite{Schenke:2012fw} have been shown - when combined with a subsequent viscous hydrodynamic evolution - to yield very successful interpretation of the measured azimutal flow distributions \cite{Gale:2012rq} and of other related observables \cite{McDonald:2016vlt}. In alternate approaches, the non-equilibrium nature of the initial stages has  been captured in several versions of effective kinetic theories based on either the Boltzmann equation  \cite{Arnold:2002zm,Xu:2004mz,Kurkela:2015qoa,Keegan:2016cpi}, or the Kadanoff-Baym equations \cite{Cassing:2008sv,Cassing:2009vt}. 

This work concentrates on the study and the analysis of the very early stages of a relativistic nuclear collision. We will make use of the Boltzmann transport equation, and use the fact that the partonic interactions are dominated by small-angle scattering and low momentum transfer, which permits a diffusion treatment in terms of a Fokker-Planck equation \cite{Lifshitz:1981xx,Blaizot:2014jna,Churchill:2020yny}. Our theoretical treatment of the non-equilibrium ensemble of quarks, antiquarks, and gluons is summarized in the next section. 
The experimental variable chosen to characterize the initial state needs to be of a penetrating nature, impervious to final state interactions.  Two obvious candidates are QCD jets \cite{[{See, for example, }][{ and references therein.}]Connors:2017ptx} and electromagnetic radiation \cite{[{See, for example, }][{ and references therein.}]Gale:2018ofa}: here we focus on the latter. 
Because the electromagnetic interaction is much weaker than the strong interaction that governs the  QGP evolution -- $\alpha/\alpha_s 
\ll 1$ -- photons and dileptons produced in heavy-ion collisions travel to the detectors essentially unscathed. By now, an impressive body of work has been devoted to the calculation and measurement of real and virtual photons, for conditions at both RHIC and the LHC \cite{[{See, for example, }][{ and references therein.}]Gale:2018ofa}, but less attention has been devoted to the electromagnetic emissivity of the pre-hydrodynamics phase. This is one of the purposes of this work: the production of real and virtual photons will be considered in a medium generally out of statistical equilibrium. 

This paper is organized as follows: the next section outlines the approach used to model the early-time evolving parton distributions, and to obtain the quark, anti-quark, and gluon phase space distribution functions. Section \ref{sec:photon_production} is devoted to details of the  calculation of real photon production rates and yields. Section  \ref{sec:dilepton_production} contains the calculation of lepton pair production. We compare with data where appropriate. We then devote Section \ref{sec:results} to a discussion of the results, and conclude.

\section{\label{sec:boltz} Partonic evolution and the Boltzmann Equation}

At the very early stages of heavy-ion collisions, the composition of the medium is dominated by gluon degrees of freedom released from the colliding nuclei. These gluons, whose dynamics initially follow nonlinear field equations~\cite{Kovner:1995ja}, are expected to occupy phase space with a probability inversely proportional to the strong coupling constant, $1/\alpha_s$, and with typical momentum of the order of the saturation scale $Q_s$. As the system expands, kinetic theory becomes applicable for gluons when  approximately $\tau\sim Q_s^{-1}$, and the evolution of gluons is then characterized by a time-dependent phase-space distribution function. 
Once gluons become 
on-shell particles and can be treated using kinetic theory, quark degrees of freedom can be introduced accordingly through QCD interactions, e.g., $gg\leftrightarrow q\bar q$. Quark production is ignored when $\tau\ll Q_s^{-1}$. 

The ingredients of kinetic theory are the phase-space distribution functions, denoted as $f_g(t,\x,\p)$ for gluons and $f_q(t,\x,\p)$ for quarks. The normalization of these functions gives rise to the number density of gluons and quarks respectively.
Owing to the charge conjugation symmetry of QCD, distribution functions of quarks and anti-quarks are not distinguished. Similarly, energy density and entropy density are well-defined quantities in  kinetic theory, in terms of integrals of $f_g(t,\x,\p)$ and $f_q(t,\x,\p)$~\cite{DeGroot:1980dk}. For later convenience, we define the longitudinal and transverse pressures as,
\begin{subequations}
\begin{eqnarray}
\label{eq:pressures}
	\P_L&=&\int\frac{d^3\mathbf{p}}{(2\pi)^3 E_p}p_z^2(\nu_g f_g + \nu_q f_q)\,, \\
	\P_T&=&\frac{1}{2}\int\frac{d^3\mathbf{p}}{(2\pi)^3 E_p}p_\perp^2(\nu_g f_g + \nu_q f_q)\, .
\end{eqnarray}
\end{subequations}
Note that we ignore the quark mass so that $E_p = \left| \vec{p} \right|$ and the energy density is related these pressures as $\epsilon=\P_L + 2\P_T$.


\subsection{The Diffusion Approximation}

The evolution of the phase-space distribution function is described by the Boltzmann equation. It is written as 
\begin{eqnarray}
\label{eq:boltz}
\frac{d}{dt}f_g(t,\x,\p) 
&=&\mathcal{C}_g[f_g(t,{\x},{\p}),f_q(t,\x,\p)]\\
\frac{d}{dt}f_q(t,\x,\p) 
&=& \mathcal{C}_q[f_g(t,{\x},{\p}),f_q(t,\x,\p)]
\end{eqnarray}
where $\mathcal{C}_g$ and $\mathcal{C}_q$ are the collision integrals, which in our current work are determined 
by $2\leftrightarrow2$ scattering processes from QCD. One may further assume that scatterings among quarks and gluons are dominated by those with small angles, so that the collision integral is simplified as a Fokker-Planck diffusion term~\cite{Lifshitz:1981xx}. In the presence of quarks, an extra source term contributes as well, thus in total one has~\cite{Blaizot:2014jna}
\begin{eqnarray}
	\mathcal{C}_g[f_g(t,{\x},{\p}),f_q(t,\x,\p)]&=& -\nabla_{\mathbf{p}}\cdot\mathcal{J}_g + \S_g \\
	\mathcal{C}_q[f_g(t,{\x},{\p}),f_q(t,\x,\p)]&=& -\nabla_{\mathbf{p}}\cdot\mathcal{J}_q + \S_q
\end{eqnarray}
where 
\begin{eqnarray}
	\mathcal{J}_g &=& -4 \pi \alpha_s^2N_c\mathcal{L}\Big[\mathcal{I}_a\nabla_{\mathbf{p}}f_g + \mathcal{I}_b\frac{\mathbf{p}}{p}f_g(1+f_g)\Big] \\ 
	\mathcal{J}_g &=& -4 \pi \alpha_s^2C_f\mathcal{L}\Big[\mathcal{I}_a\nabla_{\mathbf{p}}f_q + \mathcal{I}_b\frac{\mathbf{p}}{p}f_q(1-f_q)\Big]
\end{eqnarray}
are the effective currents, and 
\begin{eqnarray}
	\mathcal{S}_g &=& \frac{4 \pi \alpha_s^2C_FN_f\mathcal{L}\mathcal{I}_c}{p}[f_q(1+f_g)-f_g(1-f_q)]\\
	\mathcal{S}_q &=& -\frac{4 \pi \alpha_s^2C_F^2\mathcal{L}\mathcal{I}_c}{p}[f_q(1+f_g)-f_g(1-f_q)]
\end{eqnarray}
are the sources. The constant integrals in the current and source are 
\begin{eqnarray}
	\mathcal{I}_a &=& \int \frac{d^3\mathbf{p}}{(2\pi)^3}[N_c f_g(1+f_g) + N_f f_q(1-f_q)] \\
	\mathcal{I}_b &=& \int \frac{d^3\mathbf{p}}{(2\pi)^3}\frac{2}{p}(N_c f_g + N_f f_q) \\
	\mathcal{I}_c &=& \int \frac{d^3\mathbf{p}}{(2\pi)^3}\frac{1}{p}(f_g+f_q).
	\label{Ic}
\end{eqnarray}
Note that 
$\mathcal{I}_c$ effectively describes the conversion of a quark/anti-quark to a gluon, due to the exchange of a quark/anti-quark with the medium with small momentum. In the above equations, $N_c$ and $N_f$ denote the number of colors and number of flavors respectively, $C_F$ is the square of the Casimir operator of the colour $SU(N_c)$ group in the fundamental representation and is given by $C_F = (N_c^2 -1)/(2N_c)$. The Coulomb logarithm $\L$ is a divergent integral that is related to the strong coupling constant, $\L\sim \log \alpha_s^{-1}$. Alternatively, in realistic simulations the logarithm $\L$ can be taken dynamically, if one 
explicitly quantifies the UV ($q_{\rm max}$) and IR ($q_{\rm min}$) cutoffs and writes $\L=\log (q_{\rm max}/q_{\rm min})$~\cite{Blaizot:2014jna}. 

One may verify that the collision integrals conserve energy density $\e$\footnote{
In a system with one-dimensional Bjorken expansion which we consider throughout this work, both energy density and number density decay as a function of time. The conservation of energy density and number density is reflected in the evolution equation,
\begin{eqnarray}
\partial_\tau \e + \frac{\e+\P_L}{\tau}=0\,,\qquad
\partial_\tau (n\tau) =0\,.\nonumber
\end{eqnarray}
}. The total number density of quarks and gluons, i.e., $n=n_g+n_q+n_{\bar q}$, is also conserved provided that IR gluon modes do not lead to divergence, otherwise there would be a $\delta(\p)$ in the gluon distribution function corresponding to a gluon Bose-Einstein condensate (BEC)~\cite{BLAIZOT201268}. 
In the case of elastic 2-to-2 scatterings and Bjorken expansion, the gluon BEC presents as long as the initial gluon occupation exceeds some critical value. However, if the initial gluon occupation is not sufficiently large, the produced gluon BEC is transient and the system eventually approaches local thermal equilibrium. Quarks and gluons are then described using equilibrium distribution functions:
the Bose-Einstein distribution for gluons and the Fermi-Dirac distribution for quarks, with temperature $T$ and a finite effective chemical potential $\mu$. In our calculations of pre-equilibrium photons and dileptons, we initialize the system according to the realistic collisions in experiments at RHIC and the LHC (see discussions later in \Sect{sec:f0}), for which a gluon BEC is always presents during the pre-equilibrium system evolution.

The obtained equations can be made dimensionless with the help of the momentum saturation scale $Q_s$, by scaling momenta: ${\p}\rightarrow {\p}/Q_s$. For the evolution time, we take the combined factor $4\pi\alpha_s^2\L$ as a constant which can be also absorbed to define the dimensionless time parameter $t\rightarrow (4\pi\alpha_s^2\L) tQ_s$. Note that the effective strength of strong coupling constant is determined as long as this constant factor is specified. In this work we will consider $4\pi\alpha_s^2\L\sim 1$ which implies $\alpha_s\sim 0.234$. 

\subsection{Bjorken expansion of the QGP}

The early time evolution of QGP in high energy heavy-ion collisions is dominated by a longitudinal 
expansion along the collision beam axis. We describe this system using the Bjorken model, meaning that the system is boost invariant along the collision beam ($z$-axis) and translationally invariant in directions transverse to the collision beam ($x$ and $y$). The symmetry becomes apparent if one writes in terms of the proper time $\tau=\sqrt{t^2-z^2}$ and the space-time rapidity $y=\tanh^{-1}(z/t)$, such that all physical quantities depend only on $\tau$. Accordingly, this Bjorken symmetry simplifies the Boltzmann equation. The $z=0$ slice is of particular interest as $\tau\rightarrow t$ and the distribution function reduces to a function of transverse momentum $p_\perp$ and longitudinal momentum $p_z$ such that
\begin{eqnarray}
&&
\frac{d}{dt}f_g(t,{\x},{\p}) 
\rightarrow \left[\partial_t - \frac{p_z}{t}\partial_{p_z}\right]f_g(t,p_\perp,p_z)\,, \\
&&
\frac{d}{dt}f_q(t,{\x},{\p}) 
\rightarrow \left[\partial_t - \frac{p_z}{t}\partial_{p_z}\right]f_q(t,p_\perp,p_z).\hspace{1em}
\end{eqnarray}
The term proportional to $1/t$ reflects the nature of expansion along $z$, whose contribution is strong at early times.

\begin{figure}[!htb]
	\begin{centering}
		\includegraphics[width=\linewidth]{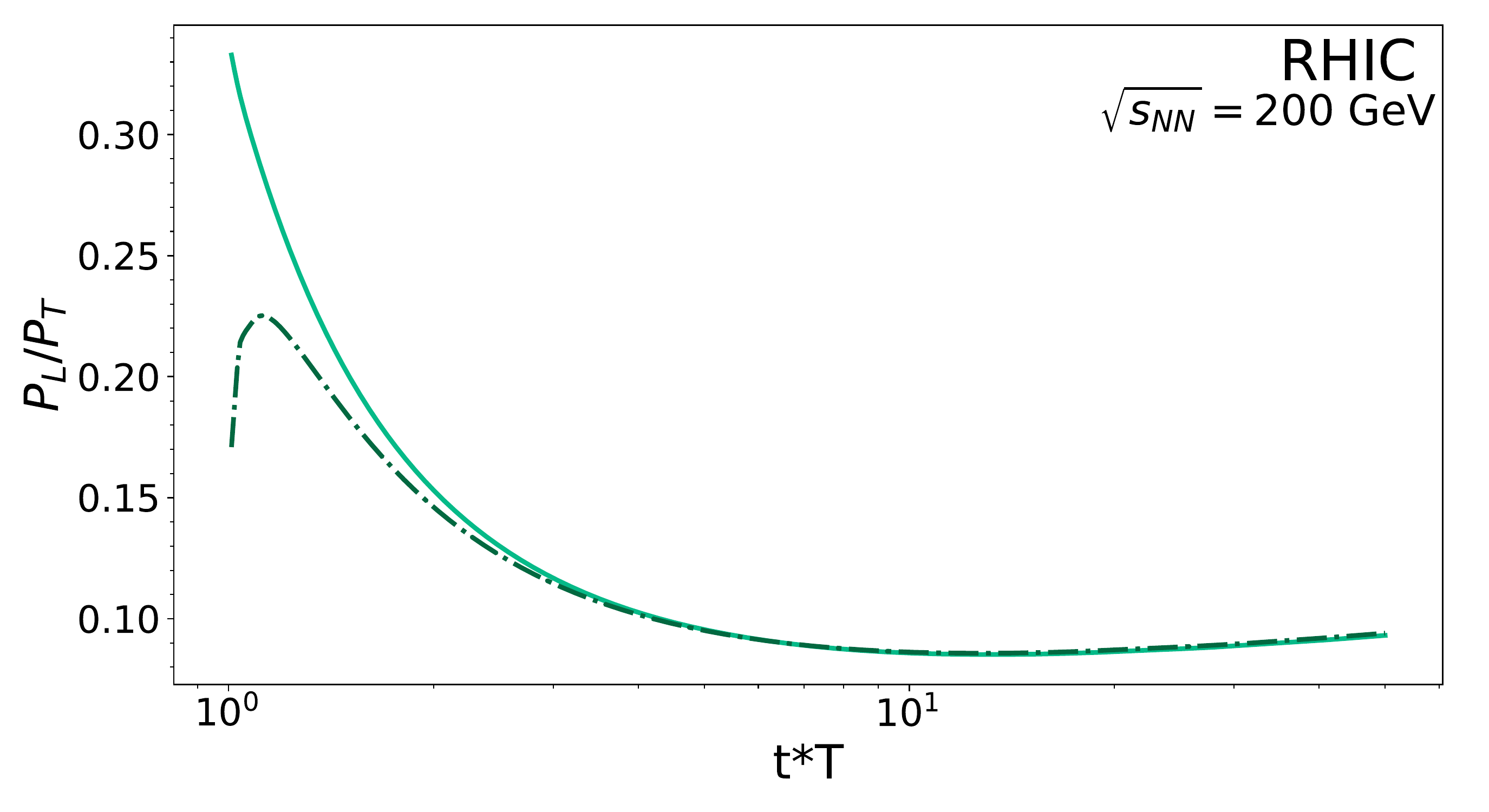}
		\includegraphics[width=\linewidth]{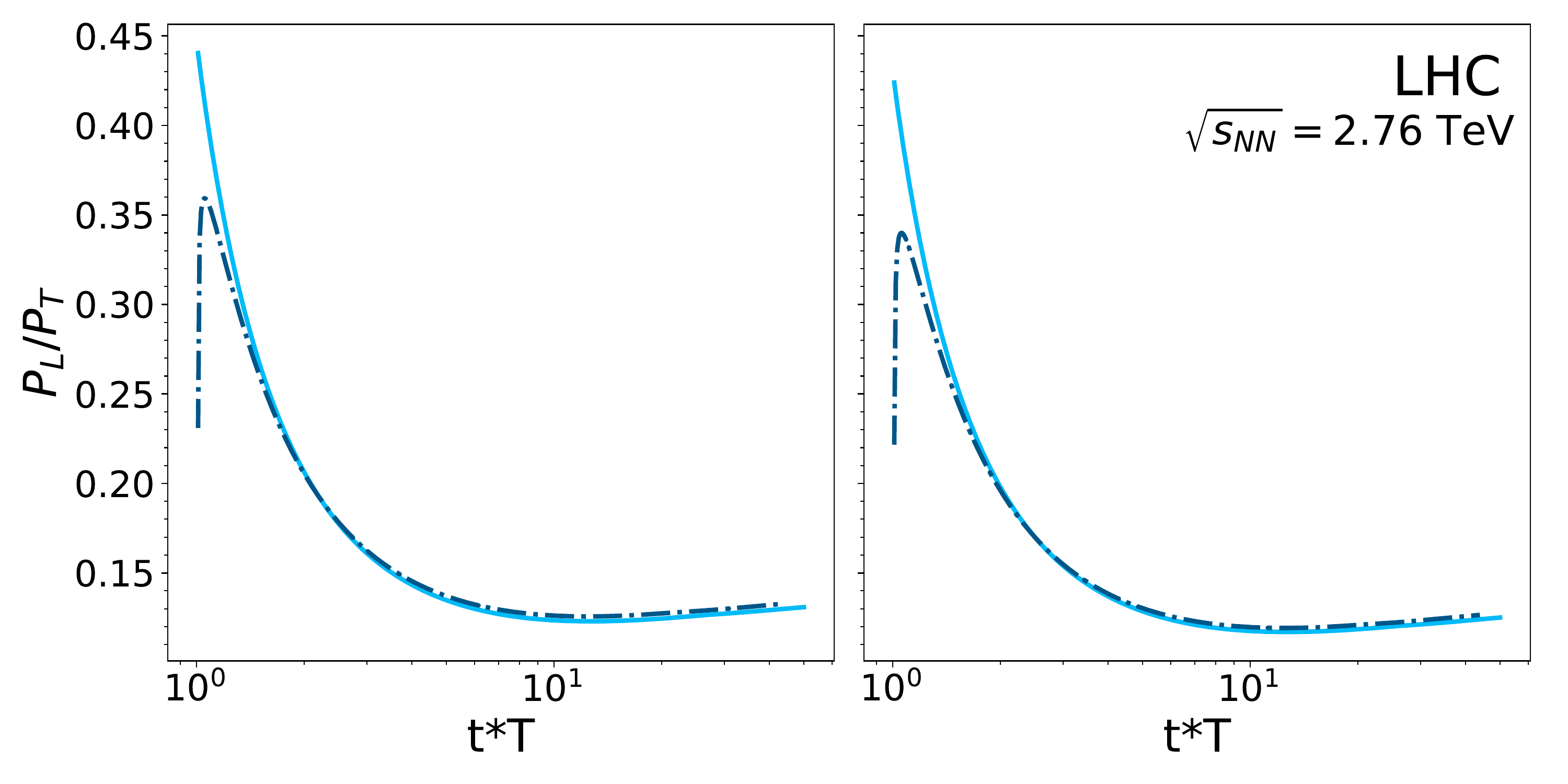}
		\includegraphics[width=\linewidth]{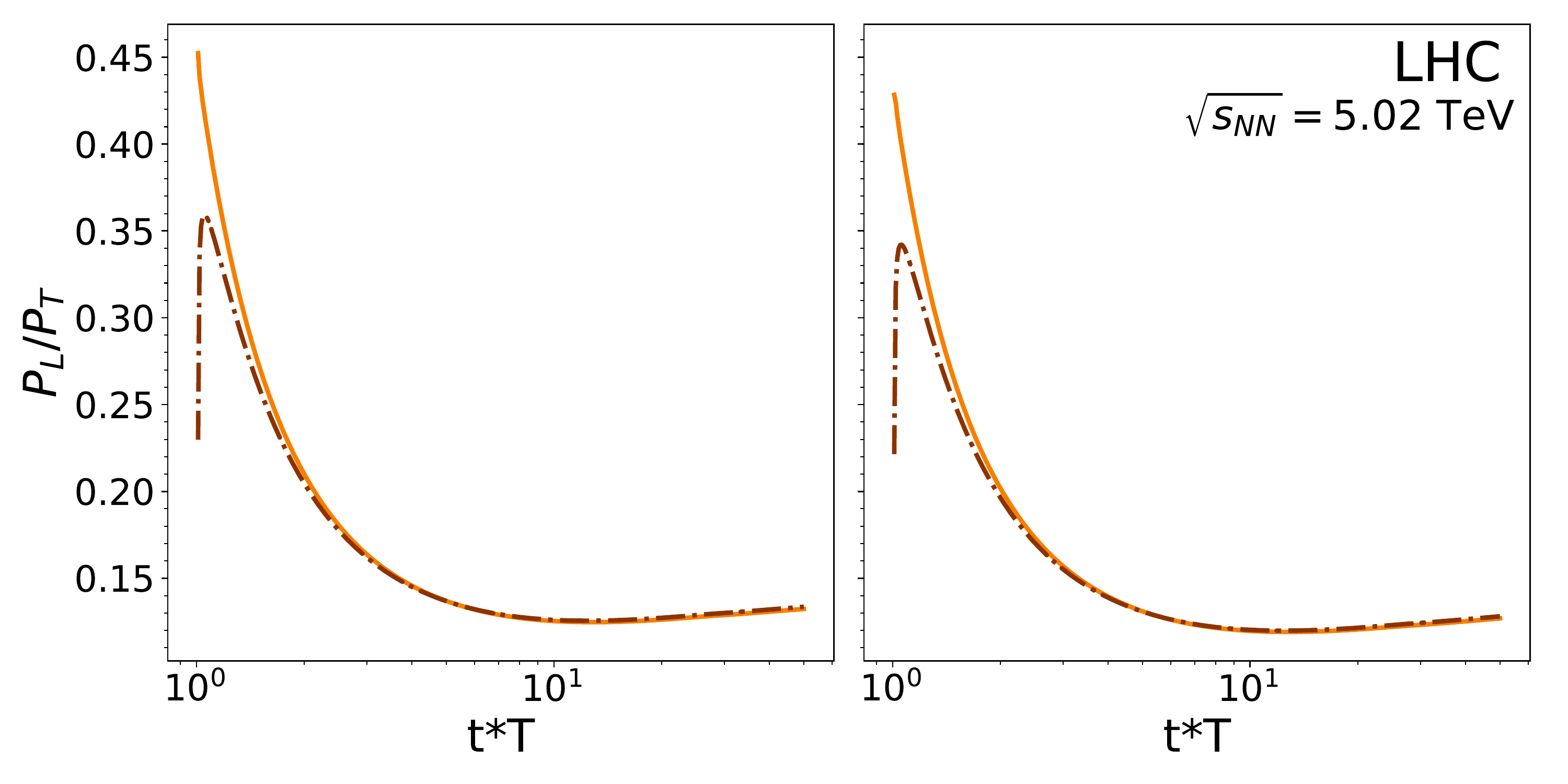}
		\caption{A study of the effect of the momentum asymmetry parameter $\xi$ on the ratio of transverse to longitudinal pressure for Au+Au collisions at RHIC (top) and Pb+Pb collisions (middle, bottom) in the 0-20\% centrality class at RHIC: $\sqrt{s}=200$ GeV with $Q_s = 1$ GeV (top panel), LHC: $\sqrt{s}=2.76$ TeV with $Q_s = 1$ GeV (middle-left), $Q_s = 2$ GeV (middle-right), LHC: $\sqrt{s}=5.02$ TeV with $Q_s = 1$ GeV (bottom-left), $Q_s = 2$ GeV (bottom-right).)}
		\label{fig:P_L/P_T}
	\end{centering}
\end{figure}

We initialize the system at $t_0 Q_s=1$ solely with gluons, as described by the  distribution function~\cite{Romatschke:2003ms}
\begin{eqnarray}\label{eq:gluonIC}
	f_g(t_0,p) = f_0\theta\Big(1-\frac{\sqrt{p_\perp^2 +  p_z^2\xi^2}}{Q_s}\Big)\,,
\end{eqnarray}
while $f_q(t_0,p)=f_{\bar q}(t_0,p)=0$. This gluon dominance  is inspired by the colour glass picture \cite{Gelis:2010nm}. The parameter $\xi$ is used to introduce an initial momentum anisotropy which leads to an initial  pressure anisotropy. Given \Eq{eq:gluonIC} and the definition of pressures in \Eq{eq:pressures}, it is not difficult to find that, $\P_L/\P_T<1$ when $\xi>1$. Pressure anisotropy is a quantity that characterizes how far an expanding system deviates from an ideal hydrodynamic description. For instance, in dissipative hydrodynamics, the difference of pressures is proportional to the viscous corrections to the stress tensor, $\P_L-\P_T\sim \eta/t$~\cite{Blaizot:2017lht}, where $\eta$ in this context is the shear viscosity. In our calculations, we consider two types of initial gluon distribution functions with $\xi=1.0$ and $\xi=1.5$, and we initialize the system evolution with a pressure
isotropy $\P_L/\P_T=1$ and a pressure anisotropy $\P_L/\P_T\approx 0.5$ respectively~\cite{Blaizot:2017lht}.
It should be emphasized that, owing to the fast longitudinal expansion, the early-stage evolution tends to 
drive the system further away from equilibrium, until collisions among quarks and gluons become dominant. 

Although, the overall out-of-equilibiurm effect is stronger in the system evolution with $\xi=1.5$ comparing to the $\xi=1.0$ case, the system evolution rapidly becomes universal. This effect of an attractor solution has been observed in the solution of kinetic theory within the relaxation time aproximation, QCD effective kinetic theory, as well as out-of-equilibrium hydrodynamics \cite{Florkowski:2017olj,Giacalone:2019ldn,Berges:2020fwq}. As a consequence of the attractor solution, out-of-equilibrium system evolution tends to merge into one single, universal path, irrespective of initial conditions. In \Fig{fig:P_L/P_T}, the pressure anisotropy $\P_L/\P_T$ is plotted as a function of $\tau T$ with respect to the nucleus-nucleus collisions carried out at RHIC and the LHC. The effective temperature $T$ is estimated via the Landau's matching condition, i.e., $\epsilon=\epsilon_{\rm eq}\propto T^4$. In all these calculations, with initial pressure anisotropy $\xi=1.0$ and 1.5, attractor behavior of the system evolution present with a universal curve realized after a short period of time, $\tau T\gtrsim 10$. Because of this attractor behavior, the dependence of final results on the switching time to hydrodynamics from kinetic theory can be suppressed. In our calculations, we use $\tau_{\rm hydro}=0.4$ fm/c as the switching time to hydrodynamics and solve the pre-equiibrium stage evolution for $Q_s^{-1}\le \tau\le 0.4$ fm/c.

\subsection{The determination of $f_0$}\label{sec:f0}

Unlike the initial state pressure anisotropy that varies as a result of quantum fluctuations in the classical gluon field evolution, the initial state gluon occupation is a fixed quantity according to the multiplicity yield in heavy-ion collisions. In \Eq{eq:gluonIC}, given a specified value of the $\xi$ parameter, the initial gluon occupation is mostly determined by the constant $f_0$ and the saturation scale $Q_s$.
We shall empirically take $Q_s=1$ GeV at the top RHIC energy, and allow $Q_s$ to vary between 1 GeV and 2 GeV at the LHC, for nucleus-nucleus collisions with $\sqrt{s_{\rm NN}}=2.76$ TeV and $5.02$ TeV~\cite{Lappi:2011gu} respectively. This freedom is used to explore the sensitivity of results obtained herein to a specific values of the saturation scale.

\begin{figure}
	\begin{centering}
		\includegraphics[width=0.9\linewidth]{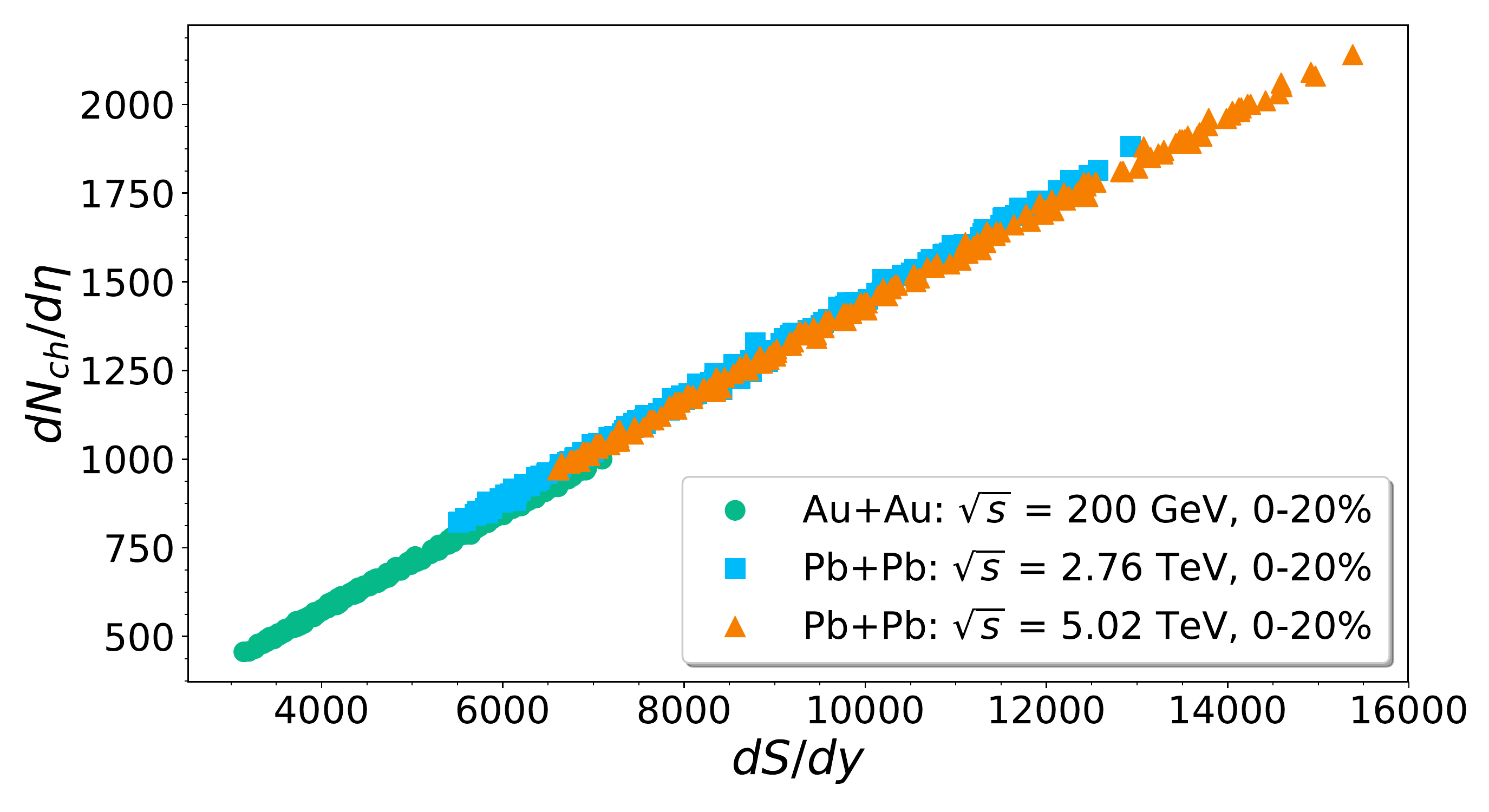}
		\caption{ Charged particle multiplicity per pseudo-rapidity as a function of initial state entropy per space-time rapidity, from event-by-event hydrodynamical simulations. Green and blue points are events pertaining to RHIC Au+Au collisions and LHC Pb+Pb collisions, respectively. The viscous fluid-dynamical simulations are performed with MUSIC  \cite{hydro,McDonald:2016vlt}.  
		}
		\label{fig:entropy_multiplicity}
	\end{centering}
\end{figure}

To determine the value of $f_0$, we will use the  relation between the initial state entropy per space-time rapidity, $dS/d y$, deduced from hydrodynamical simulations and the empirical charge particle multiplicity per pseudo-rapidity\footnote{The pseudo-rapidity is traditionally written as $\eta$, and so is the shear viscosity. The appropriate meaning should be clear in context.}
\begin{equation}\label{eqn:entropy_calc}
\frac{dS}{d y } \sim 7.14 \frac{dN_{\rm ch}}{d \eta}\,,
\end{equation}
as seen in Fig. \ref{fig:entropy_multiplicity}. 
The initial state entropy density is calculated using details of the time evolution of hydro calculations that correctly reproduce final state observables \cite{hydro,McDonald:2016vlt}. 
Note that the initial time of hydrodynamical evolution would be the final time of the pre-equilibrium evolution solved from the Boltzmann equation, i.e., $\tau_{\rm hydro}=0.4$ fm/c. The manifestly linear relation simply follows from the idea that entropy per particle yield in heavy-ion collisions depends little on rapidity~\cite{Bjorken:1982qr,Gubser:2008pc}, and the fact that entropy production during the hydrodynamical evolution is subdominant. The constant 7.14 contains the information on the effective microscopic degrees of freedom in the QCD equation of state~\cite{Bazavov:2014pvz}. It can be extracted from event-by-event hydrodynamical simulations. Results are shown in \Fig{fig:entropy_multiplicity} for Au+Au collisions at RHIC and Pb+Pb collisions at the LHC in the centrality class 0-20\%. These results are obtained using by now standard hydrodynamical simulations of heavy-ion collisions~\cite{McDonald:2016vlt},  with $\tau_{\rm hydro}=0.4$ fm/c, and $\eta/s=$0.12, for Au+Au collisions  at $\sqrt{s_{\rm NN}}=200$ GeV, and $\eta/s = 0.13$ Pb+Pb at $\sqrt{s_{\rm NN}}=2.76$ and 5.02 TeV. A temperature dependent $\zeta/s$ is used \cite{Ryu:2015vwa}. The observed experimental flow harmonics can be well reproduced, together with other key observables. 

As shown in \Fig{fig:entropy_multiplicity}, the linear relation between the charged particle multiplicity and initial entropy is indeed apparent, although the slopes from Au+Au and Pb+Pb are very slightly different. Note that the extracted constant is smaller comparing to the simple estimate using a conformal EoS, which is approximately 7.5~\cite{Gubser:2008pc}. The  \Eq{eqn:entropy_calc} allows one to work out the appropriate entropy in the beginning of hydrodynamical simulations, given experimental results for charged particle multiplicities and the  access to the details of the hydrodynamical simulations. Values corresponding to the measured charged particle multiplicity at RHIC~\cite{Adler:2004zn} and the LHC~\cite{Chatrchyan:2012mb,Adam:2015ptt} are shown in \Tab{tab:f0_values}.

\begin{table}
	\centering
	\begin{tabular}{c|c|c|c|c|c|c}
		& \begin{tabular}[c]{@{}c@{}}$\mathbf{\sqrt{s_{NN}}}$\\ {[TeV]}\end{tabular} & $\mathbf{dS/d y}$     & \begin{tabular}[c]{@{}c@{}}$\mathbf{A_T}$\\ {[fm$^2$]}\end{tabular} & \begin{tabular}[c]{@{}c@{}}$\mathbf{Q_s}$\\ {[GeV]}\end{tabular} & \begin{tabular}[c]{@{}c@{}}$\mathbf{f_0}$\\ ($\xi$ = 1.0)\end{tabular} & \begin{tabular}[c]{@{}c@{}}$\mathbf{f_0}$\\ ($\xi$ = 1.5)\end{tabular} \\ \hline 
		\textbf{RHIC}                 & 0.20                                                            & 5000                   & 100.58                                                     & 1.0                                                     & 2.25                                                          & 3.81                                                          \\ \hline
		\multirow{2}{*}{\textbf{LHC}} & \multirow{2}{*}{2.76}                                           & \multirow{2}{*}{13700} & \multirow{2}{*}{124.25}&1.0                                                    & 6.65                                                          & 11.10                                                         \\
		&                                                                 &                      &  & 2.0                                                     & 5.75                                                          & 9.50                                                          \\ \hline
		\multirow{2}{*}{\textbf{LHC}} & \multirow{2}{*}{5.02}                                           & \multirow{2}{*}{14500} &\multirow{2}{*}{127.75} &1.0                                                     & 7.00                                                          & 11.75                                                         \\
		&                                                                 &                        & &2.0                                                     & 6.00                                                          & 10.25                                                        
	\end{tabular}
	\caption{Initial gluon population ($f_0$) values determined by matching hydro initial conditions to experimental observables, for a $0 - 20\%$ centrality class (note that the energy and $Q_s$ entries are centrality-independent). See main text for details. 
	}
	\label{tab:f0_values}
\end{table}

The dominant entropy production is from the pre-equilibrium stage of the system evolution. We solve the Boltzmann equation with respect to initial condition \Eq{eq:gluonIC}, up to $\tau_{\rm hydro}=0.4$ fm/c. Entropy density is a well-defined quantity in the kinetic theory. For quarks and gluons, one has
\begin{align}
&s_{g}\equiv -\nu_g\int \frac{d^3\p}{(2\pi)^3}\left[ f_g \log f_g - (1+f_g)\log(1+f_g) \right],\\
&s_{q}\equiv - \nu_q \int \frac{d^3\p}{(2\pi)^3}\left[ f_q \log f_q  + (1- f_q)\log(1- f_q)\right]\,,    
\end{align}
which gives the total entropy per space-time rapidity as
\be
\frac{dS}{d y } = \tau A_T (s_g+s_q)\,.
\ee
In realistic calculations, the transverse overlapping area $A_T$ can be determined effectively, although there is not a  ``standard'' way to calculate the nuclear overlap area. Since our study pertains to early time dynamics, the geometry associated with the Glauber model is appropriate. Overlap areas calculated using Glauber Monte-Carlo were calculated for different systems colliding at different energies, and binned in different centrality classes. These values were tabulated in Ref. \cite{Loizides:2017ack}, and they are written symbolically here as $A_T$. 

\begin{figure}[!htb]
	\begin{centering}
		\includegraphics[width=\linewidth]{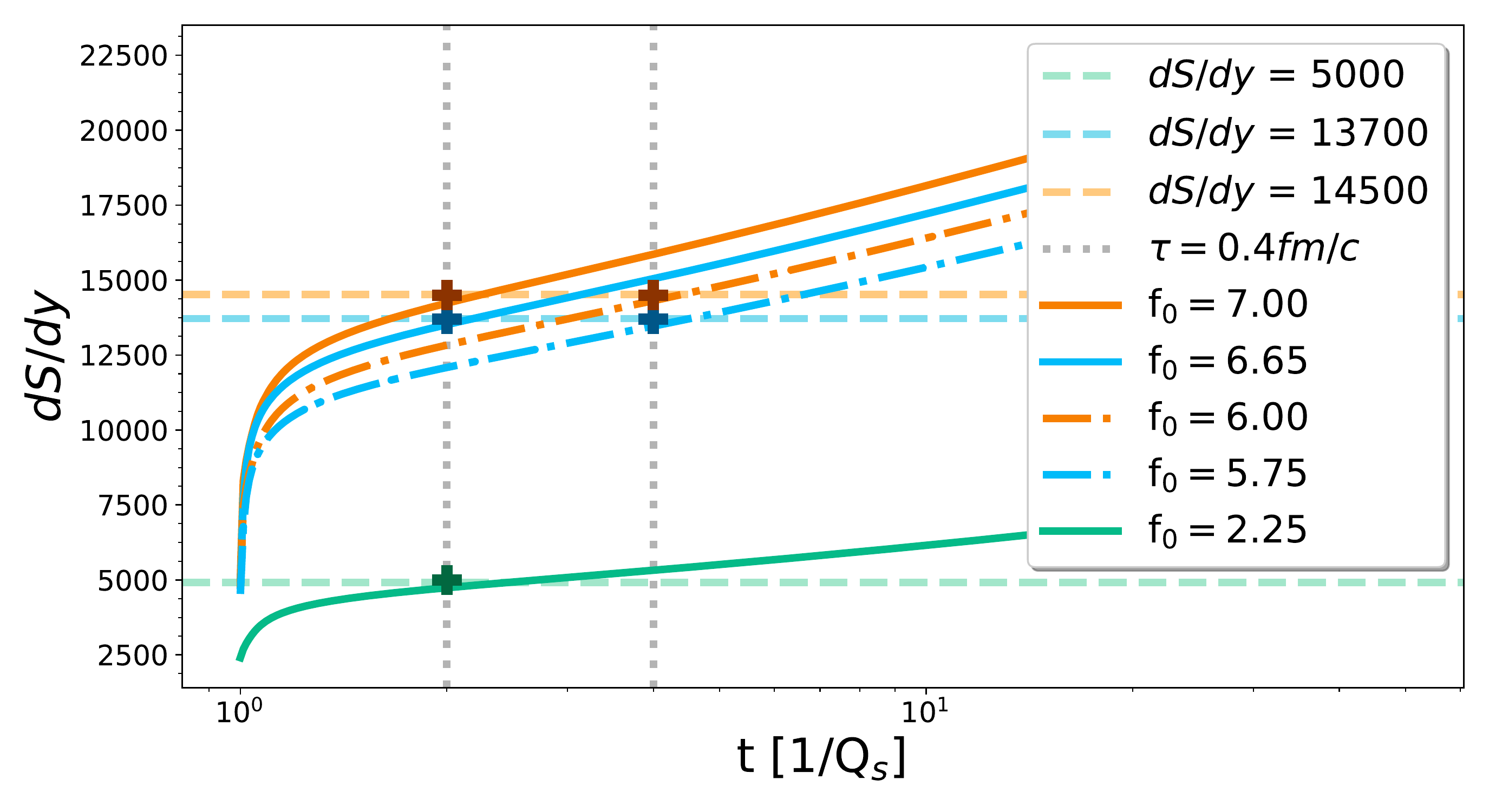}
		\includegraphics[width=\linewidth]{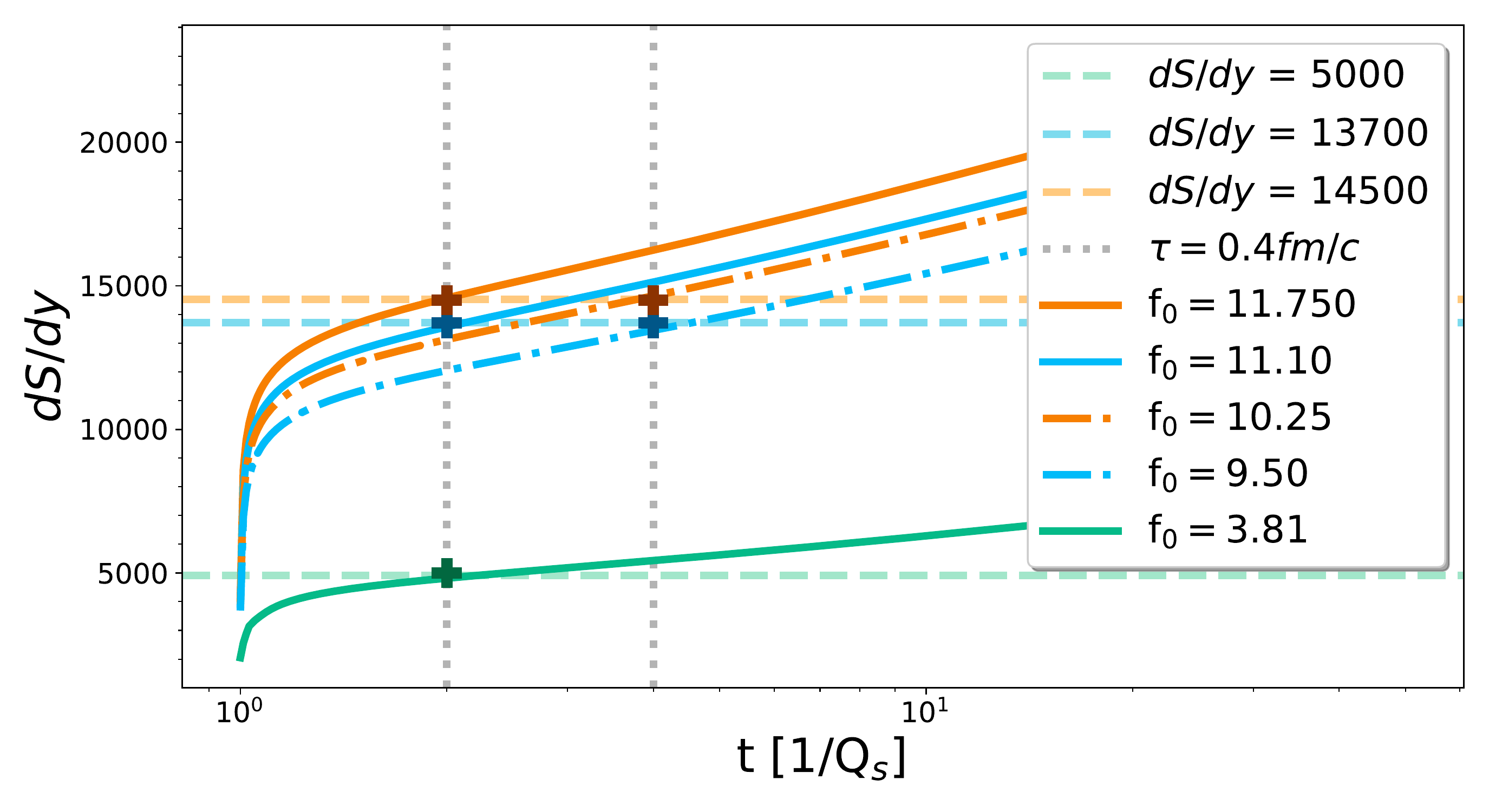}
		\caption{Entropy evolution for various energies used to determine the appropriate $f_0$ value for each system, where $\xi$ = 1.0 (top) or 1.5 (bottom). Grey vertical lines correspond to t = 0.4 fm/c for $Q_s$ = 1 and 2 GeV, coloured horizontal lines correspond to $dS/d y$ from Table \ref{tab:f0_values}. Quantities related to RHIC (200 GeV) are shown in green, LHC (2.76 TeV) are shown in blue, and LHC (5.02 TeV) are shown in orange. Results are summarized in Table \ref{tab:f0_values}; see the main text for data references.}
		\label{fig:entropy}
	\end{centering}
\end{figure}

In principle, the pre-equilibrium entropy production monotonously depends on the values of $f_0$, hence by tuning $f_0$, one is able to match the desired entropy per space-time rapidity, with respect to the realistic colliding systems. \Fig{fig:entropy} illustrates the evolution of entropy in the pre-equilibrium stage and the matching procedure using various values of $f_0$ in our calculations. We now have a partonic evolution model which can be used to calculate the emission of electromagnetic radiation for the duration of the non-equilibrium phase.

\section{Photon Production}
\label{sec:photon_production}
\begin{figure}[!htbp]
	\begin{centering}
		\includegraphics[width=0.8\linewidth]{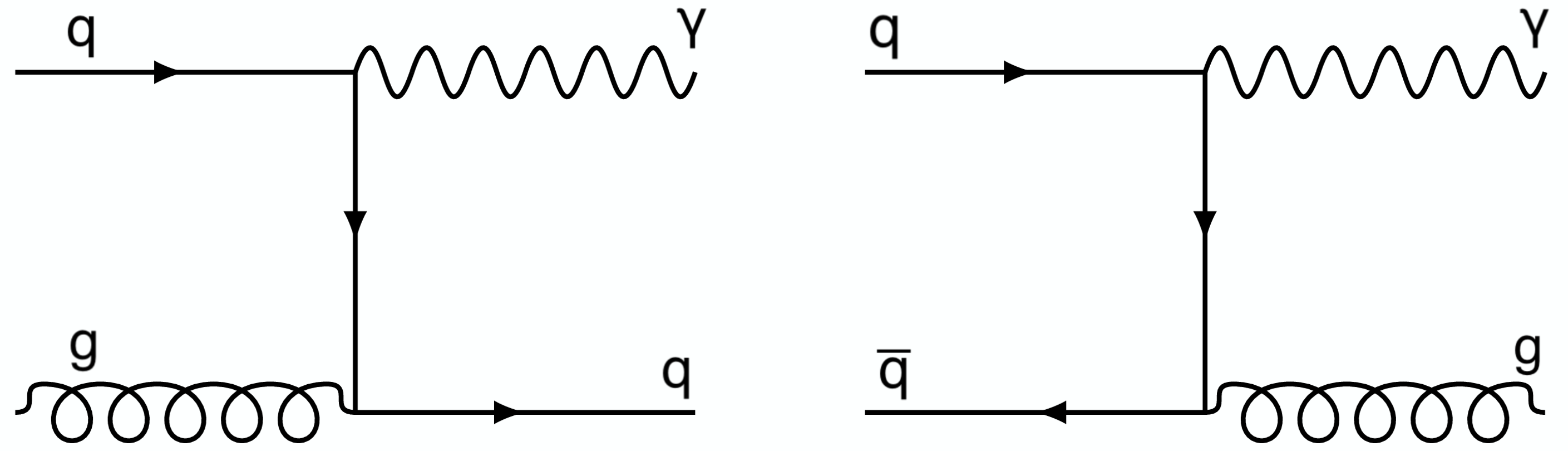}
		\caption{Photon production through Compton scattering (left) and quark-antiquark annihilation (right). The appropriate $s$ and $u$ channels are not shown, but included in the calculation.}
		\label{fig:photon}
	\end{centering}
\end{figure}

To reiterate, electromagnetic probes are penetrating and  relay information complementary to that contained in strongly interacting ones. For studies of pre-equilibrium dynamics their value is matched only by jets in the hard sector, and unparalleled in the soft sector. First, we investigate the production of real photons.  In a system comprised of only quarks and gluons, both dileptons and photons can be produced through the annihilation of a quark with an anti-quark. However, unlike dileptons, photons canal be so generated through the Compton scattering process, in which a quark or anti-quark scatters with a gluon. Feynman diagrams for both those channels are shown in Figure \ref{fig:photon}.  As was the case for the virtual photon emission treated earlier \cite{Aurenche:2002wq}, the real photons attributed to the LPM effect are not considered explicitly here. That contribution can be comparable in magnitude to the sum of Compton and quark antiquark annihilations, depending on the energy of the emitted photon \cite{Arnold:2001ms}. 

\subsection{Out-of-Equilibrium Photon Emission Rate} 

The production rate of photons can be derived starting with the expression for the production of on-shell photons, 
\begin{eqnarray}
E\frac{d^3R}{d^3p} &=& 
\sum_{i} \int \frac{d^3p_1}{(2\pi)^32E_1}\frac{d^3p_2}{(2\pi)^32E_2}\frac{d^3p_3}{(2\pi)^32E_3} \frac{1}{2(2\pi)^3}\nonumber\\
&\hspace{1em}& \times |\mathcal{M}_i|^2(2\pi)^4 \delta^4(P_1+P_2-P_3-P) \nonumber\\
&\hspace{1em}& \times f_1(\mathbf{p}_1)f_2(\mathbf{p}_2)[1 \pm f_3(\mathbf{p}_3)],
\end{eqnarray}
where the degeneracy factor 
has been absorbed into the amplitude $|\mathcal{M}_i|^2$. Summation over $i$ denotes contributions from the Compton and quark-antiquark annihilation channels. The distribution functions $f_1$, $f_2$ and $f_3$ are for quarks and gluons, corresponding to the scattering processes respectively.

Following the procedure outlined, for example,  in Ref. \cite{Berges:2017eom}, the expression for the off-equilibrium photon production rate can be derived using the small-angle approximation which assumes the dominance of low momentum transfer between scattering particles. 
One may thus expand the kinematic variables in terms of the exchanged momentum $\mathbf{q}=\mathbf{p}-\mathbf{p_1}$, and perform the kinematic integrals.
Writing explicitly the gluon and fermion distributions, $f_g$ and $f_q (=f_{\bar q})$, one arrives at expressions for the photon production rate \cite{Berges:2017eom}. 
The  rate from the annihilation process is given by
\begin{eqnarray}
E\frac{d^3R}{d^3p} &=& \frac{40\alpha\alpha_s}{9\pi^2} \mathcal{L'} f_q(\mathbf{p})\int \frac{d^3p'}{(2\pi)^3}\frac{1}{p'}f_q(\mathbf{p'}) [1 + f_g(\mathbf{p'})]. \hspace{2em}
\end{eqnarray}
For the Compton scattering contribution, a similar derivation can be performed which yields the expression 
\begin{eqnarray}
E\frac{d^3R}{d^3p} &=& \frac{40\alpha\alpha_s}{9\pi^2} \mathcal{L'} f_q(\mathbf{p})\int \frac{d^3p'}{(2\pi)^3}\frac{1}{p'}f_g(\mathbf{p'})[1 - f_q(\mathbf{p'})], \hspace{2em}
\end{eqnarray}
where the logarithmic divergence is given by 
\begin{eqnarray}\label{eq:L}
\mathcal{L'} = \int_{\Lambda_{IR}}^{\Lambda_{UV}} \frac{dq}{q} = \ln\frac{\Lambda_{UV}}{\Lambda_{IR}}.
\end{eqnarray}
The IR cutoff is given by the the Debye mass scale $m_D \sim g T$ and the UV is regulated by the temperature $T$. 

Summing the Compton and annihilation contributions gives the expression 
\begin{eqnarray}
E\frac{d^3R}{d^3p} &=& \frac{40\alpha\alpha_s}{9\pi^2} \mathcal{L'} f_q(\mathbf{p})\int \frac{d^3p'}{(2\pi)^3}\frac{1}{p'}[f_g(\mathbf{p'}) + f_q(\mathbf{p'})] \hspace{2em} \\ \nonumber
&=& \frac{40\alpha\alpha_s}{9\pi^2} \mathcal{L'} f_q(\mathbf{p}) \, \mathcal{I}_c
\label{eq:net_rate}
\end{eqnarray}
where $\mathcal{I}_c$ is defined in Eq. (\ref{Ic}). 
The expansion of the production rate in terms of the exchanged momentum  described previously simplifies the analytic expressions, but avoids the details of the HTL (Hard Thermal Loops) regulation of IR divergences \cite{Kapusta:1991qp}.  To fix the overall scale of the net rates, the quark and gluon distribution functions, $f_q$ and $f_g$, in Eq. (\ref{eq:net_rate}) are replaced by thermal distribution functions. 
The factor $\mathcal L'$ is treated as an adjustable constant fixed by matching the result of this expression to that of the equivalent analytical expression from \cite{Kapusta:1991qp}. This procedure yields a constant of $\sim {\mathcal{O}} (1)$. Importantly, this will neglect the contribution associated with the Landau-Pomeranchuk-Migdal (LPM) effect, which has been shown to contribute an approximate additional factor of 2 \cite{Arnold:2001ms,Ghiglieri:2016tvj} in equilibrium. However, owing to phase space considerations, the non-equilibrium dynamics  may well have a different effect on the LPM contribution than it does on $2 \to 2$ processes. We  consider this possible factor to be part of the systematic theoretical uncertainties in this study, and we argue that leaving it out in fact provides a conservative yield estimate.  An appropriate assessment of complete leading order non-equilibrium photon production in the context of the dynamical approach used here requires the numerical implementation of a  field-theoretical analysis \cite{Hauksson:2017udm} which we leave for future work. This  would include a non-equilibrium assessment of Debye screening and LPM effects. 


\subsection{Out-of-Equilibrium Photon Yield}
To calculate the off-equilibrium photon yield, the photon production rate is converted using 
\begin{eqnarray}\label{eq:photon_yield_conversion}
E\frac{d^3R}{d^3p} = E\frac{dN}{d^4Xd^3p} = \frac{dN}{\tau d\tau d^2\mathbf{x_{\perp}} dy dy_p d^2\mathbf{p_{\perp}}}. 
\end{eqnarray}
This means that an integral over $y $ and $\tau$ is needed in order to obtain an expression of the form $dN/dy_p d^2\mathbf{p_{\perp}}$. 

In the photon production rate calculation, it was assumed that $y = 0$. Therefore, the $y$ dependence needs to be restored for non-zero values of $y$. To do this, a change of variables from $p_z(y)=\tilde{p}_z=p_{\perp}\sinh(y_p-y)$ is performed. This can be further rewritten knowing that $p_z=p_{\perp}\sinh(y_p)$, so that change of variables becomes 
\begin{eqnarray}\label{eq:pz_tilde}
\tilde{p}_z=p_{\perp}\sinh(\sinh^{-1}(p_z/p_{\perp})-y)
\end{eqnarray}
such that $p_z(0)=p_z$ returns the original equation. Thus, upon performing this change of variables and noting that the distribution functions are also a function of $\tau$, an integration over $y$ and $\tau$ gives
\begin{eqnarray}
\frac{dN}{d^2\mathbf{x_{\perp}} dy_p d^2\mathbf{p_{\perp}}} &=& \frac{16}{3\pi^2}\alpha\alpha_s \mathcal{L'} \int \tau d\tau d y f_q(p_{\perp}, \tilde{p}_z, \tau) \mathcal{I}_c . \hspace{2em}
\end{eqnarray}
The integration over $x_{\perp}$ yields $A_T$, the overlapping transverse area of the two colliding nuclei. 
The expression for the off-equilibrium photon yield can therefore be written as 
\begin{eqnarray}
\frac{dN}{dy_p d^2\mathbf{p_{\perp}}} &=& \frac{16 A_T }{3\pi^2}\alpha\alpha_s \mathcal{L'} \int \tau d\tau dy f_q(p_{\perp}, \tilde{p}_z, \tau) \mathcal{I}_c.
\label{eq:photon_yield}
\end{eqnarray}
In the expression for the yield, the parton distribution functions are obtained from the time-dependent numerical solution of the Boltzmann equation, using the procedures and methods discussed in Section \ref{sec:boltz}. Eq. (\ref{eq:photon_yield}) is then integrated from an initial time of $1/Q_s$, where $Q_s$ is either 1 or 2 GeV. For the choice of a final time, we rely on analyses of heavy-ion phenomenology, as commonly practiced. Specifically, many fluid-dynamical simulations of relativistic nuclear collisions adopt a starting time of $\tau_{\rm hydro} = 0.4$ fm/c \cite{Gale:2013da}. Therefore, in this study the pre-equilibrium phase exists for a time $\tau$, with $1/Q_s \leq \tau \leq 0.4$ fm/c. 

\begin{figure}[!htb]
	\begin{centering}%
		\includegraphics[width=\linewidth]{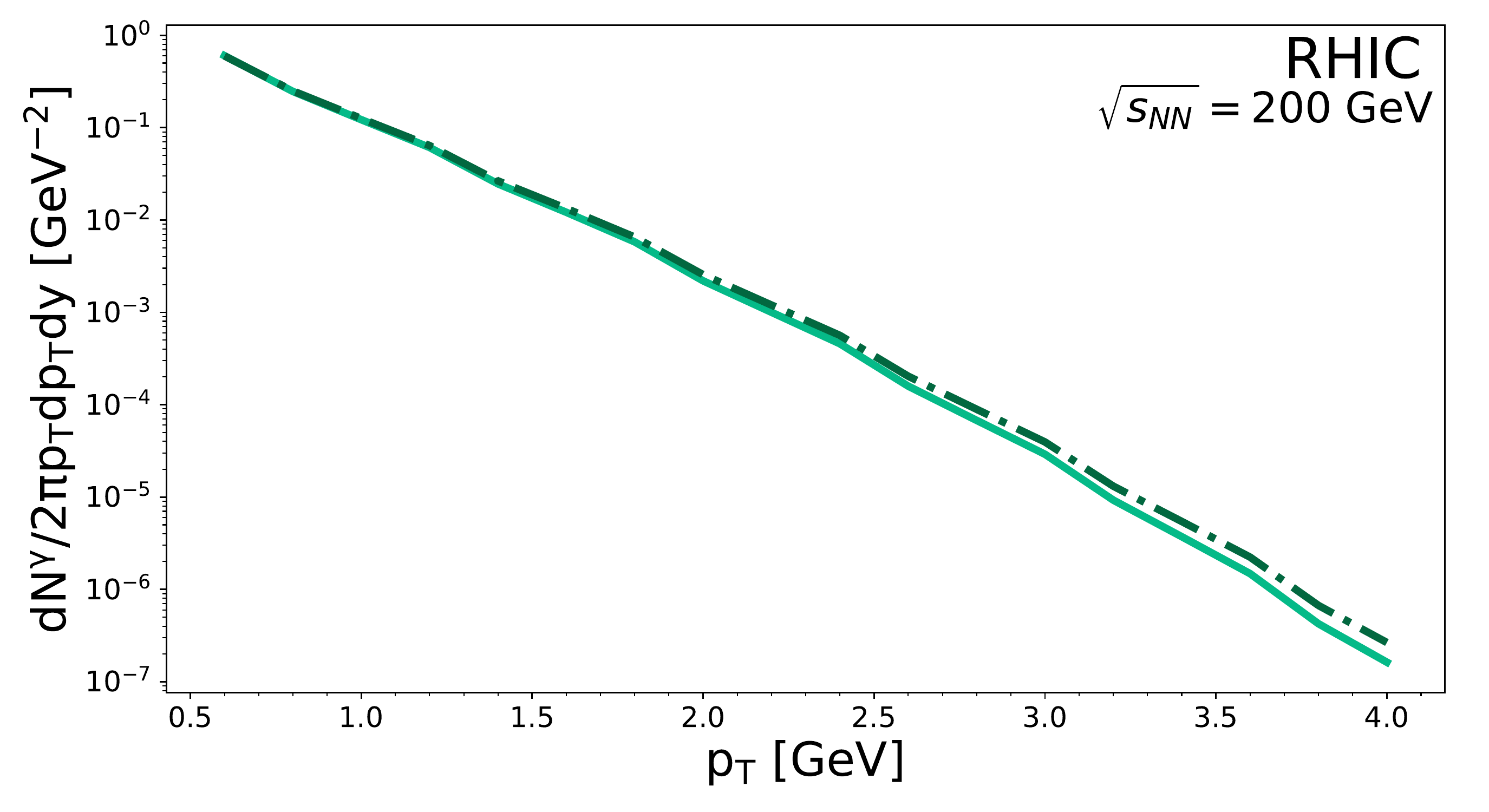}
		\includegraphics[width=\linewidth]{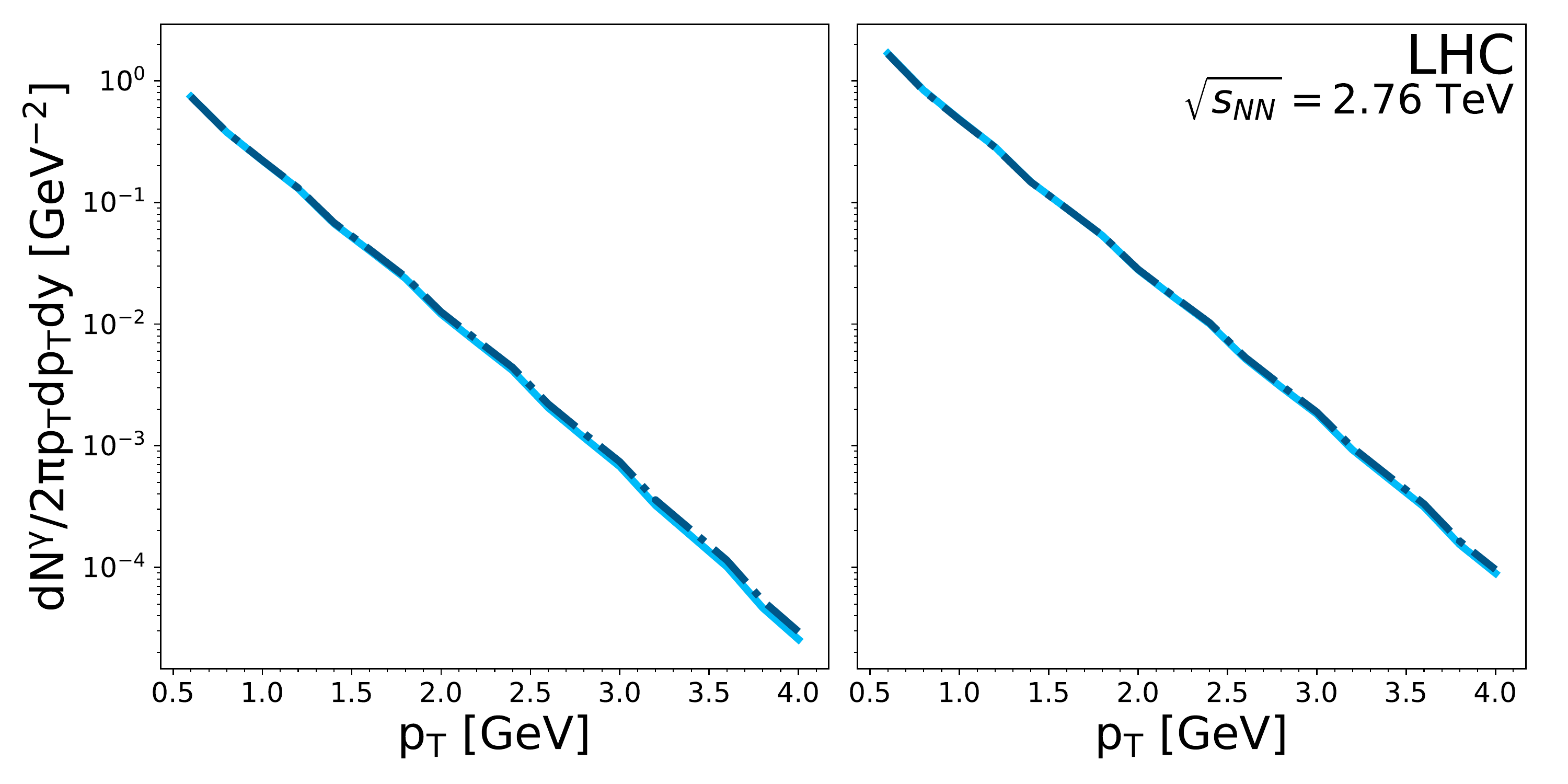}
		\includegraphics[width=\linewidth]{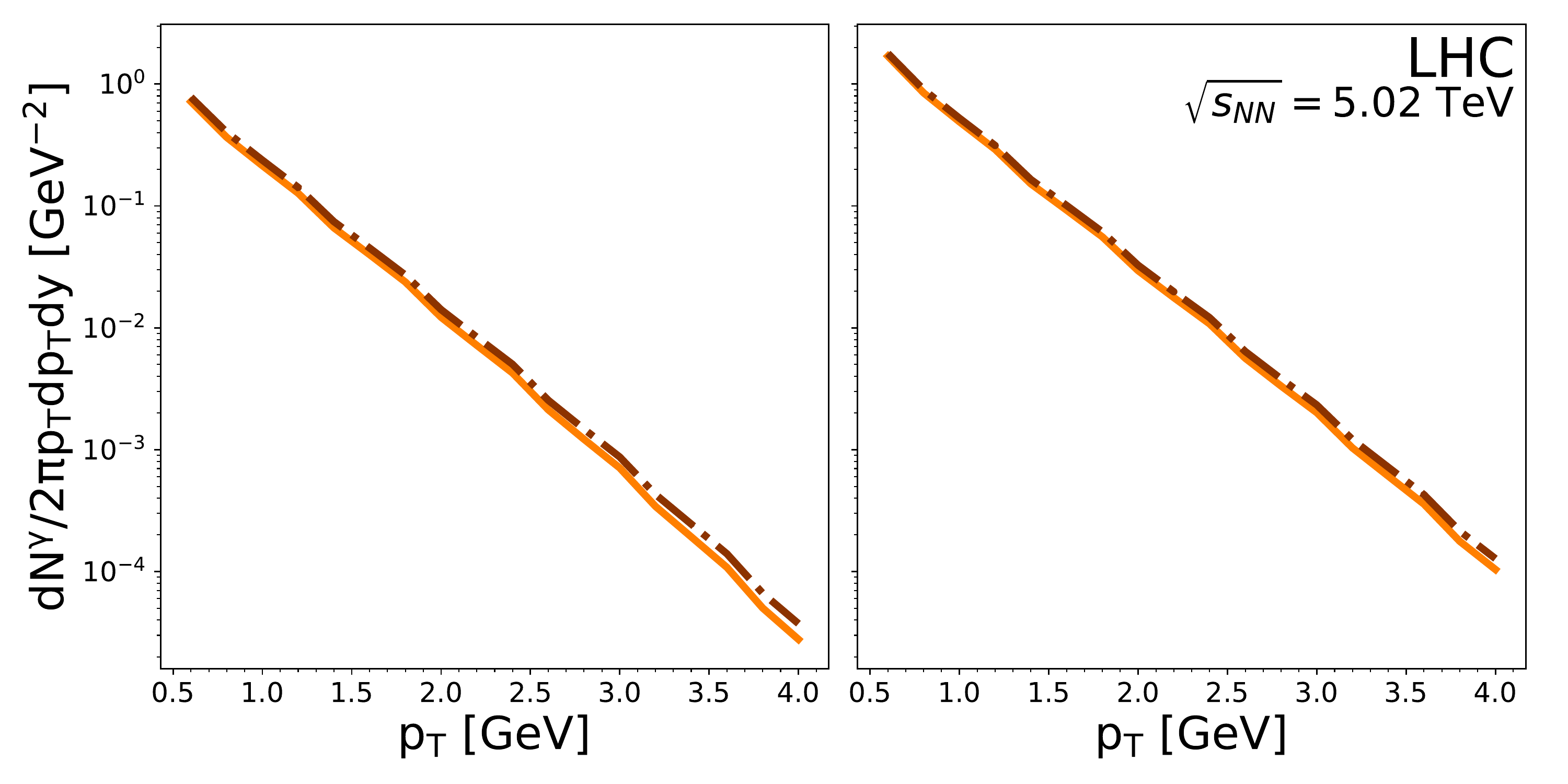}
		\caption{Off-equilibrium photon yield integrated from 1/$Q_s$-0.4 fm/c for Au+Au collisions (top) and Pb+Pb collisions (middle, bottom) at 0-20\% centrality. The plot shows the real photon spectra calculated at RHIC: $\sqrt{s}=200$ GeV with $Q_s = 1$ GeV (top panel), LHC: $\sqrt{s}=2.76$ TeV with $Q_s = 1$ GeV (middle-left), $Q_s = 2$ GeV (middle-right), LHC: $\sqrt{s}=5.02$ TeV with $Q_s = 1$ GeV (bottom-left), $Q_s = 2$ GeV (bottom-right).}
		\label{fig:photon_xi}
	\end{centering}
\end{figure}

As a first step, we study the effect of the anisotropy of the initial gluon distribution on the pre-equilibrium photon spectra. Figure \ref{fig:photon_xi} shows the result of integrating the non-equilibrium photon rates (rates evaluated with distributions from the transport calculation) from a time of $1/Q_s$ ($\sim 0.2$ fm/c for RHIC energies, and $\sim 0.1$ fm/c for LHC) to 0.4 fm/c, the end of the pre-equilibrium phase. The effect of using the  different values of the asymmetry parameter $\xi$  is shown, and its values are chosen to span the parameter space and sample a of ratios of longitudinal to transverse pressure, as defined by Eqs. (\ref{eq:pressures}). Recall that the values $f_0$ follow from requiring the multiplicity per unit pseudo-rapidity to match experimentally measured values, via a fluid-dynamical analysis. One observes that the effect of the initial gluon asymmetry on the final photon spectra is modest. Quantitatively, going from $\xi =1$ to $\xi=1.5$ we report an increase of $\sim$35\% at RHIC ($\sqrt{s_{NN}}$ = 200 GeV and $Q_s$ = 1 GeV), 
and of $\sim$5\% and $\sim$15\%  at the LHC (for $Q_s =$ 2 GeV, and $\sqrt{s_{NN}}$ = 2.76 TeV and  5.02 TeV, respectively). These numbers are for a photon transverse momentum of $p_T =$ 2.5 GeV. 
\subsection{Other sources and  experimental data}

Electromagnetic radiation is emitted throughout the entire space-time history of the collision process. Therefore, the radiation from the pre-equilibrium sources will compete with others, and will constitute only one of the contributions to the total yield. In what concerns real photons, those other  contributors are primordial nucleon-nucleon collisions. The photon production there can be calculated using next-to-leading order (NLO) perturbative QCD \cite{Aurenche:2006vj}. These photons are often referred to as ``prompt photons''. In addition, the strongly interacting medium which is modeled by viscous hydrodynamics will shine throughout its existence \cite{Paquet:2015lta}. The photons are often referred to as ``thermal photons''. Finally, the late stages, where matter falls out of thermal equilibrium can also generate photons. Those contributions must be evaluated using a transport approach \cite{Linnyk:2015rco,Schafer:2019edr}. 

\begin{figure}[!htb]
	\begin{centering}
		\includegraphics[width=\linewidth]{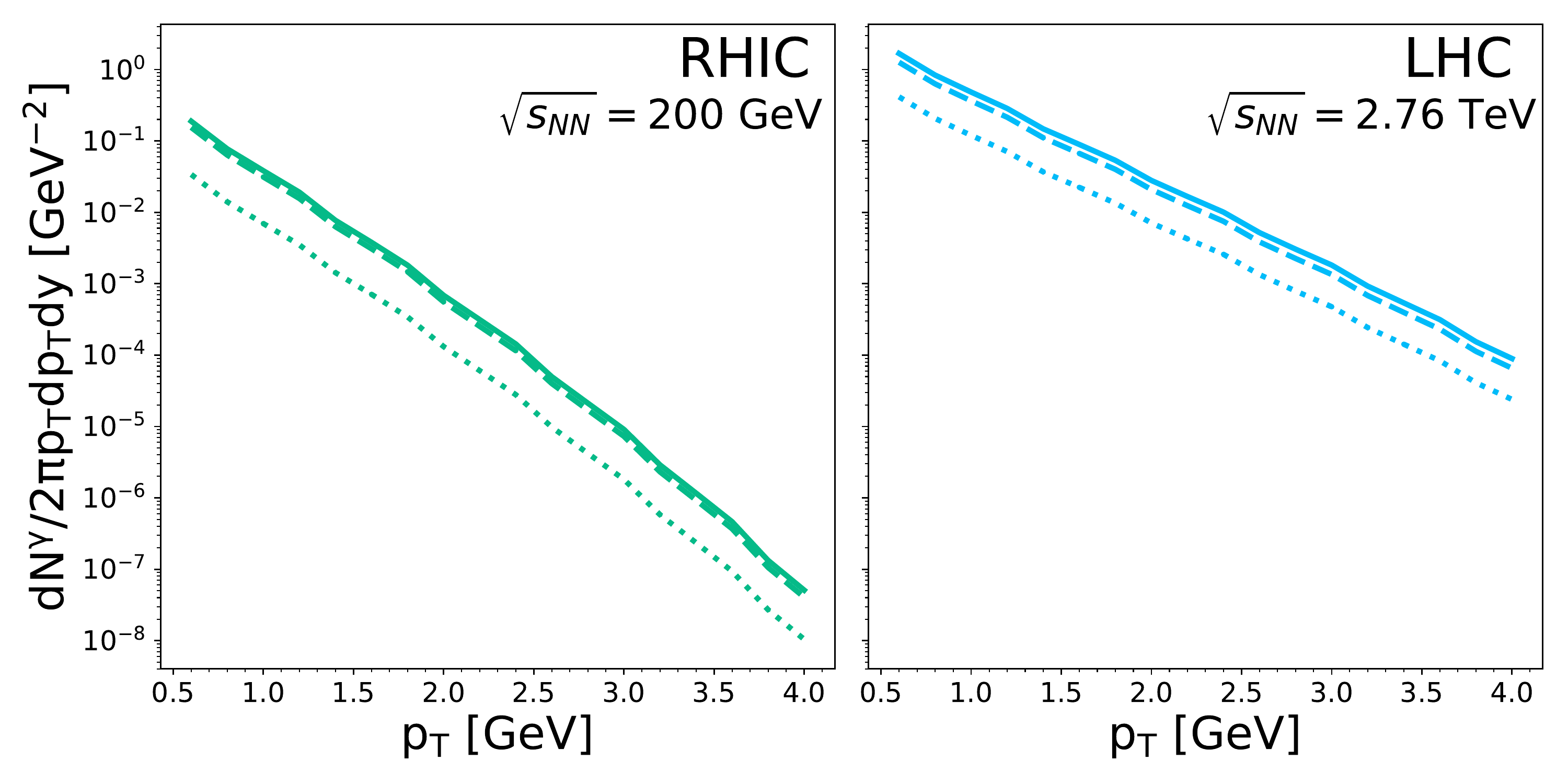}
		\caption{The total pre-equilibrium photon yield (solid line), and the individual Compton (dashed line) and quark-antiquark (dotted line) contributions. The left panel shows results of collisions performed under conditions appropriate for RHIC ($f_0 = 2.25$); the right panel shows results appropriate for the LHC ($f_0 = 5.75$). Results show here are for a $0 - 20\%$ centrality class.}
		\label{fig:prequ_Compton_qq}
	\end{centering}
\end{figure}
\begin{figure}[!htb]
	\begin{centering}
		\includegraphics[width=\linewidth]{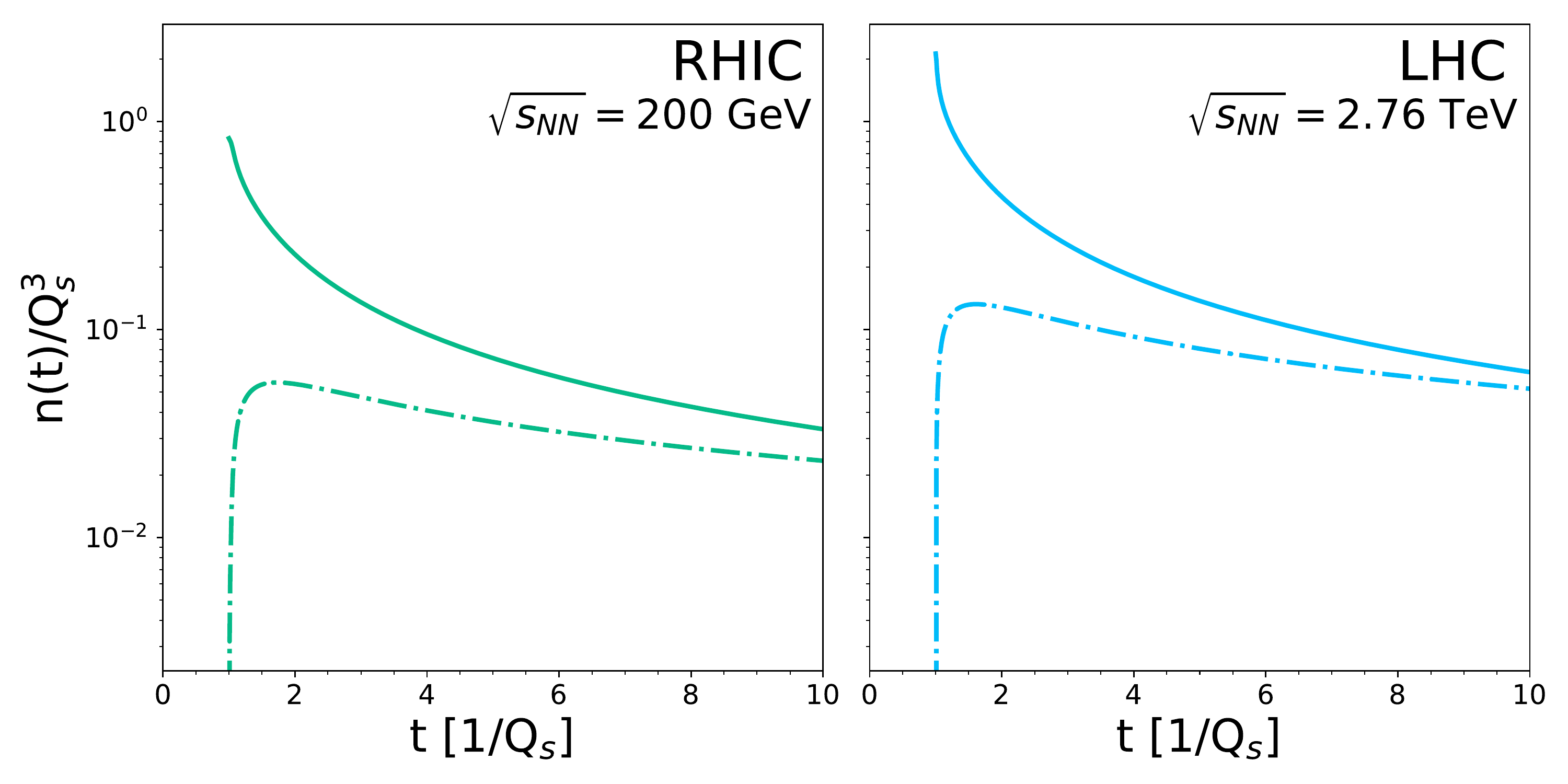} 
		\caption{The evolution in time of the gluon and fermion number density, for conditions relevant for RHIC (left panel), and for the LHC (right panel). The solid line is for gluons, the dashed-dotted line is for quarks.}
		\label{fig:number_density}
	\end{centering}
\end{figure}

Fig. \ref{fig:prequ_Compton_qq} shows the net pre-equilibrium photon yield, with the Compton and $q\bar{q}$ annihilation channels shown separately, for conditions prevalent at RHIC and at the LHC. In both plots, a striking feature is the dominance of the Compton channel over the fermion annihilation channel. This fact is easily understood in terms of the parton dynamics at work here. The Compton channel is linear in the fermionic density, whereas annihilation is quadratic. Initially, the fermions are absent, as the initial state is gluon-dominated; the quark and anti-quark populations then proceed to grow dynamically. This is illustrated in Fig. \ref{fig:number_density} which shows the time evolution of the gluonic and fermionic parton density.  The difference in intensity between the Compton and annihilation channels is therefore a direct consequence of this asymmetry in partonic content. To illustrate this point even more vividly, recall that the photon-producing Compton and quark/anti-quark annihilation rates are identical in equilibrium \cite{Wong:1995jf}. The time-evolution of an equilibrated medium would therefore populate the photon final state spectrum with an equal number of ``Compton photons'' and of ``$q \bar{q}$ photons''. Thus, pre-equilibrium photons offer unique insight into the dynamics which control the early-time chemistry of the parton population. 

\begin{figure}[!htb]
	\begin{centering}
		\includegraphics[width=\linewidth]{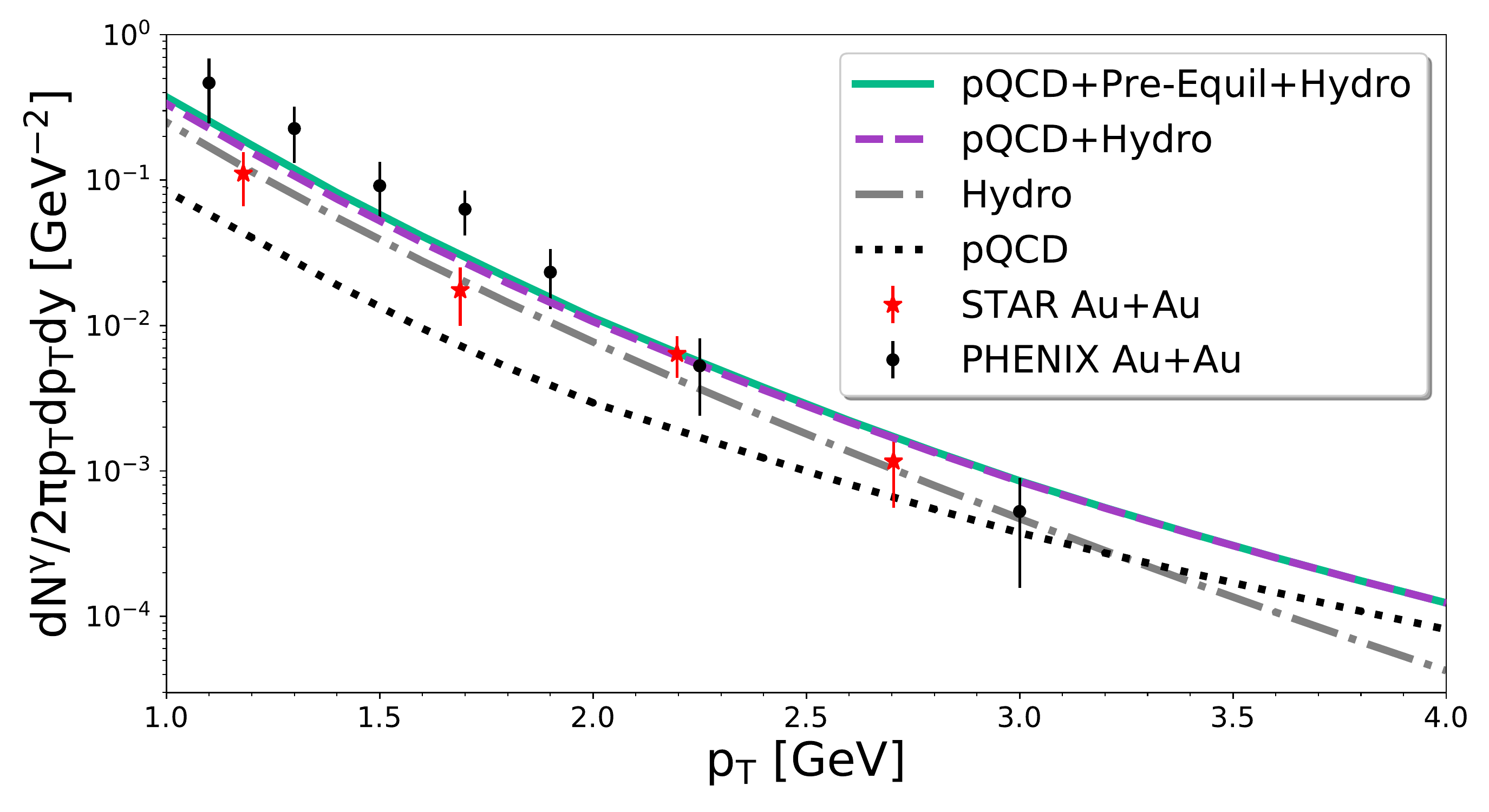}
		\caption{The sum of pQCD, pre-equilibrium, and hydro photons (see text for details) obtained in  Au+Au collisions in the 0-20\% centrality class at RHIC compared to data from STAR \cite{STAR:2016use} and PHENIX \cite{Adare:2014fwh}. Here, $Q_s$ = 1 GeV.}
		\label{fig:pQCD+prequ_+hydro_RHIC}
	\end{centering}
\end{figure}

We now turn to other sources and also consider experimental data, in order to set the scale of the early-time photon radiation. 
Figure \ref{fig:pQCD+prequ_+hydro_RHIC} shows a variety of sources, together with direct\footnote{The decay photons - those radiated by late-stage unstable hadrons - have been subtracted away in the experimental analyses.} photon data from two different experimental RHIC collaborations. The dotted line represents the photons originating from primordial nucleon-nucleon collisions, the calculation of which considers NLO pQCD contributions, together with corrections accounting for isospin and nuclear in-medium effects \cite{Paquet}. Adding to those the photons generated in the hydrodynamic phase \cite{Paquet:2015lta} (shown by the dot-dashed line) of the nuclear reaction yields the short-dashed contribution. It is important to specify that the ``hydro photons'' are corrected for viscous effects, in both the shear and bulk sector, as completely as is currently known \cite{Paquet:2015lta}.  Note that, since the late-stage electromagnetic emission models are still under development \cite{Schafer:2019edr}, those photons are simulated by letting the hydrodynamic modeling operate until $T \sim 105$ MeV \cite{Paquet:2015lta}. 

\begin{figure}[!htb]
	\begin{centering}
		\includegraphics[width=\linewidth]{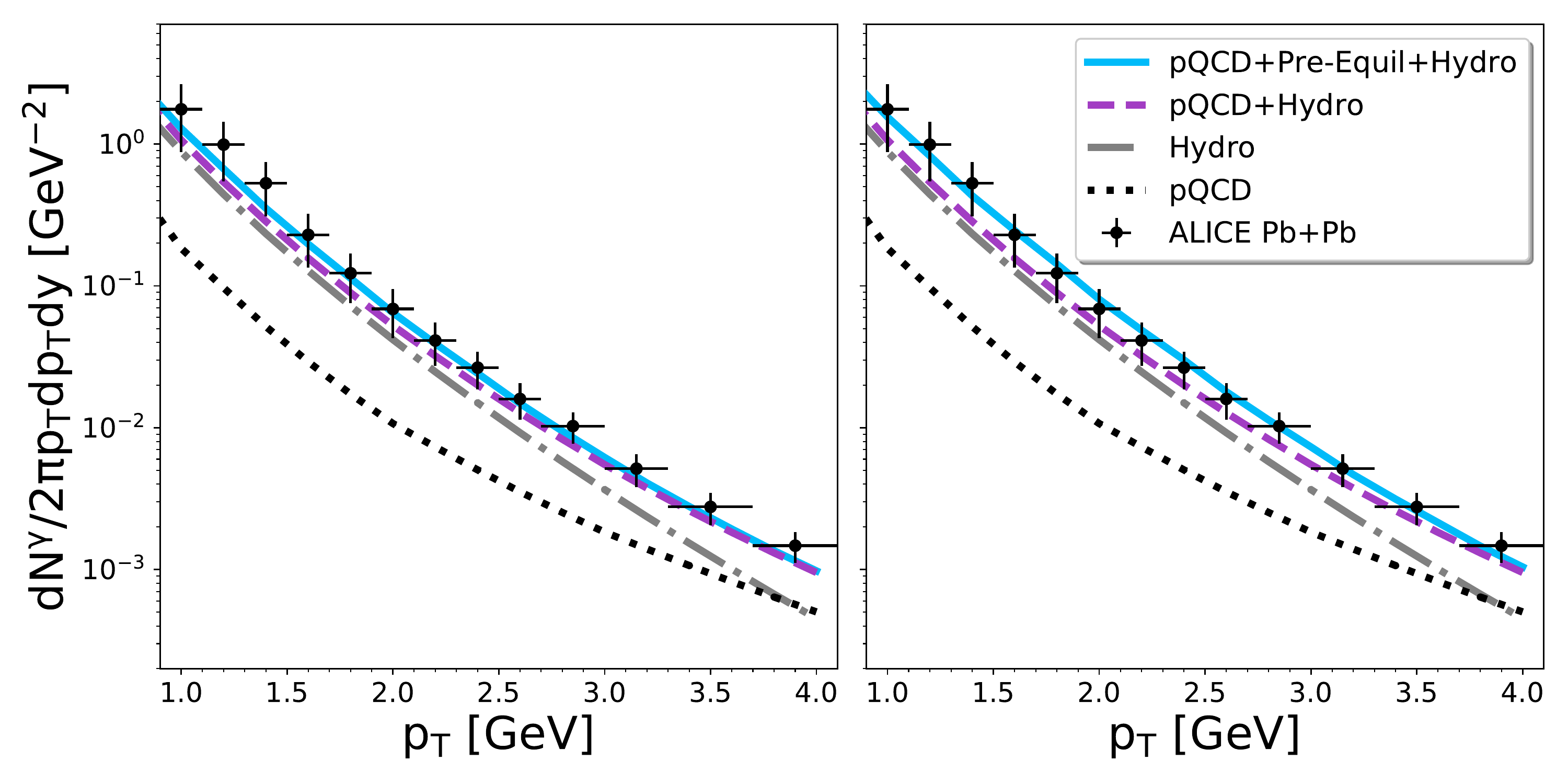}
		\caption{The sum of pQCD, pre-equilibrium, and hydro photons (see text for details) obtained in Pb+Pb collisions in the 0-20\% centrality class at the LHC at 2.76 TeV compared to data from ALICE \cite{Adam:2015lda}. Here, $Q_s$ = 1 GeV in the left panel and $Q_s$= 2 GeV on the right.}
		\label{fig:pQCD+prequ_+hydro_LHC_2p76}
	\end{centering}
\end{figure}

The solid line is the sum of all photon sources considered in this study. The difference between that line and the short-dashed one represents  the pre-equilibrium contribution. 
At a transverse momentum of $p_T = 2$ GeV, the pre-equilibrium photons represent $\sim 6\%$ of the total yield, and less than $3\%$ at $p_T =2.5$ GeV. The signature of the pre-equilibrium dynamics go down with increasing momentum, to approximately disappear at $p_T \sim 3$ GeV. The pre-equilibrium photons are also calculated for LHC conditions, and are plotted in Fig. \ref{fig:pQCD+prequ_+hydro_LHC_2p76}. 
The different sources here are as described for RHIC. In addition, a study of the effect of varying $Q_s$ is performed: the left panel uses $Q_s = 1$ GeV, whereas the right panel uses $Q_s = 2$ GeV. There, one observes a more significant $\sim 38 \%$ contribution from the pre-equilibrium photons to the total signal, at $p_T = 2.5$ GeV. The current data does not have the resolution to exclude either $Q_s$  value, and is statistically consistent with both. However, this simple model does generate a pre-equilibrium photon yield which could be within reach of contemporary experiments, provided an upgraded low-$p_T$ resolution. 

\begin{figure}[!htb]
	\begin{centering}
		\includegraphics[width=\linewidth]{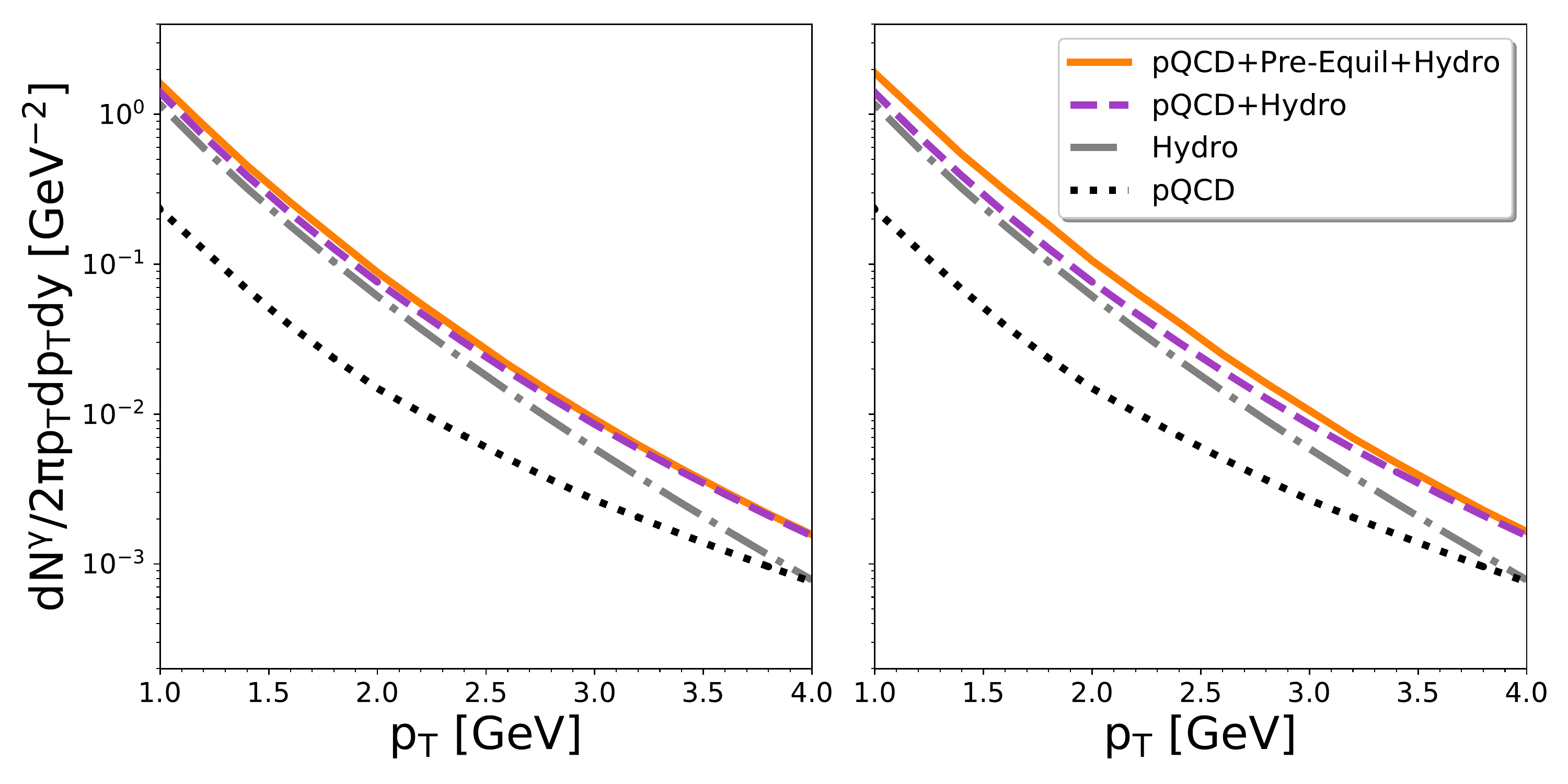}
		\caption{The sum of pQCD photons of pre-equilibrium photons (see text for details) obtained in Pb+Pb collisions in the 0-20\% centrality class at the LHC at 5.02 TeV. Here, $Q_s$ = 1 GeV in the left panel and $Q_s$= 2 GeV on the right. }
		\label{fig:pQCD+prequ_+hydro_LHC_5p02}
	\end{centering}
\end{figure}

The  study performed in this work can be extended to include photon spectra to be measured and analyzed  at the LHC, featuring Pb+Pb collisions at an energy of $\sqrt{s} = 5.02$ A TeV. This prediction appears in Fig. \ref{fig:pQCD+prequ_+hydro_LHC_5p02}. 
 At $p_T = 2.5$ GeV, pre-equilibrium photons represent $\sim 25\%$ of the total photon yield, and its effect can be seen to persist up to $p_T \sim 4$ GeV. The pre-equilibrium component shines about as brightly at both LHC energies, but the contribution from the hydro phase increases with increasing energy, therefore outshining the pre-equilibrium contribution more than at the lower colliding energy.

\section{\label{sec:dilepton_production}Dilepton Production}
The dense system of quarks and gluons formed immediately after relativistic heavy-ion collisions can also be studied using virtual photons, or dileptons produced through quark/anti-quark annihilation. This section contains the details of the derivation. 


\subsection{Out-of-Equilibrium Dilepton Rate}

\begin{figure}[!htbp]
	\begin{centering}
		\includegraphics[width=0.65\linewidth]{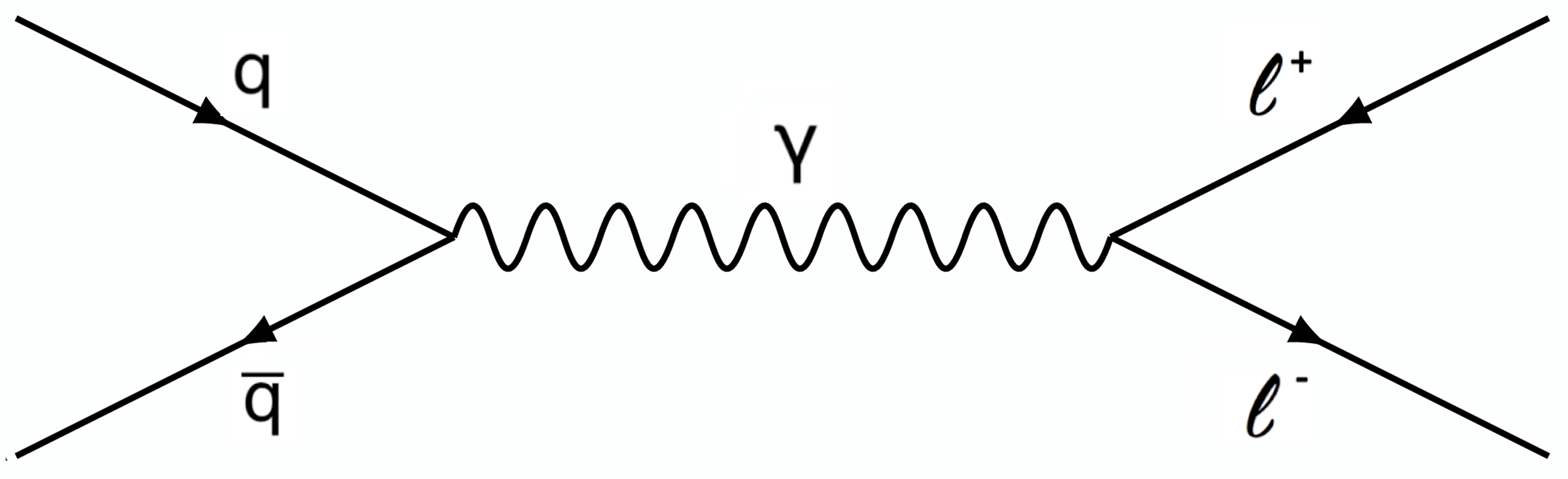}
		\caption{Dilepton production through quark anti-quark annihilation. }
		\label{fig:dilepton}
	\end{centering}
\end{figure}

From relativistic kinetic theory, the rate of production of dileptons from $q\bar{q} \rightarrow l^+l^-$ (Fig.  \ref{fig:dilepton}), can be derived \cite{Kajantie:1986dh,Gale:1987ki,Martinez:2008di,Ryblewski:2015hea}

\begin{eqnarray}\label{eq:dilepton_dR/d4Q}
\frac{dR}{d^4Q} &=& \int \frac{d^3\mathbf{p}_1}{(2\pi)^3}\frac{d^3\mathbf{p}_2}{(2\pi)^3} f(\mathbf{p}_1)f(\mathbf{p}_2)v_{q\bar{q}}\sigma_{q\bar{q}}(M) \\ \nonumber
&\hspace{1em}& \times \delta^{(4)}(Q-P_1-P_2),
\end{eqnarray}
which is the number of dileptons produced per space-time volume and four dimensional momentum-space volume. In this equation, the relativistic relative velocity is
\begin{eqnarray}
\nu_{q\bar{q}} = \frac{\sqrt{(p_1\cdot p_2)^2-m_q^4}}{E_1 E_2} = \frac{M^2}{2}
\end{eqnarray}
and the total cross section is given by 
\begin{eqnarray}
\sigma_{q\bar{q}} = F_q \tilde{\sigma}(M)
\end{eqnarray}
where 
\begin{eqnarray}
F_q = \Big[N_c(2s+1)^2\sum_{f}e_f^2\Big] 
\end{eqnarray}
\begin{eqnarray}
\tilde{\sigma}(M) &=& \frac{4\pi}{3} \frac{\alpha_{EM}^2}{M^2}\Big(1+\frac{2m_l^2}{M^2}\Big)\Big(1-\frac{4m_l^2}{M^2}\Big)^{1/2}.
\end{eqnarray}
Only $u$, $d$, and $s$ massless quarks are used and it is assumed that the rest mass of the leptons is much less than M, the centre-of-mass energy and the dilepton invariant mass. As this is the off-equilibrium case, thermal distribution functions cannot be used. Therefore, the distribution functions must come from the out-of-equilibrium solution to the Boltzmann equation \cite{Blaizot:2014jna}, where, as mentioned, the resulting distribution functions are given in terms of $p_{\perp}$, $p_{z}$, and $\tau$. 
After integrating over $\mathbf{p_2}$ and $\phi_1$, the equation becomes
\begin{eqnarray}
\frac{dR}{d^4Q} &=& \frac{\alpha_{EM}^2}{12\pi^5} \int \frac{dp_{1\perp}dp_{1z} 2p_{1\perp}f_q(p_{1\perp},p_{1z},\tau )}{E_1\sqrt{4Q_{\perp}^2p_{1\perp}^2-(2E E_1-M^2)^2}} \\ \nonumber
&\hspace{1em}& \times  f_{\bar{q}}(\sqrt{Q_{\perp}^2+p_{1\perp}^2-2E E_1+M^2},-p_{1z},\tau )
\end{eqnarray}
where 
$\bar{p}_{2\perp} = \sqrt{Q_{\perp}^2+p_{1\perp}^2-2Q_{\perp}p_{1\perp}\cos\phi_1}$ and $\bar{p}_{2z}=Q_z-p_{2z}$, such that the integral is no longer dependent on $\mathbf{p_2}$. The known identities
\begin{eqnarray}
E &=& \sqrt{M^2+Q_{\perp}^2+Q_z^2} = \sqrt{M^2+Q_{\perp}^2}\cosh(y)\\
Q_z &=& \sqrt{M^2+Q_{\perp}^2}\sinh(y)
\end{eqnarray}
were used for the boost-invariant case, where the rapidity may be set to $y=0$ as in the Bjorken model such that
\begin{eqnarray}
E &=& \sqrt{M^2+Q_{\perp}^2} \\
Q_z &=& 0.
\end{eqnarray}

The off-equilibrium dilepton production rate could also be written in terms of mass distribution knowing that 
\begin{eqnarray}
\frac{dR}{d^4Q} = \frac{dR}{MdMdyd^2Q_{\perp}},
\end{eqnarray}
where $MdM = \frac{1}{2}dM^2$ and, as before, $y=0$. Therefore, by integrating over $Q_{\perp}$, an alternate expression for the rate is given by 
\begin{eqnarray}
\frac{dR}{dM^2} &=& \frac{\alpha_{EM}^2}{12\pi^5} \int \frac{2p_{1\perp}Q_{\perp}dp_{1\perp}dp_{1z}dQ_{\perp}}{E_1\sqrt{4Q_{\perp}^2p_{1\perp}^2-(2E E_1-M^2)^2}} \\ \nonumber
&\hspace{1em}& \times f_q(p_{1\perp},p_{1z},\tau ) \\ \nonumber
&\hspace{1em}& \times f_{\bar{q}}(\sqrt{Q_{\perp}^2+p_{1\perp}^2-(2E E_1-M^2)},-p_{1z},\tau ).
\end{eqnarray}

\subsection{Out-of-Equilibrium Dilepton Yield}

The above expression can be converted into the dilepton yield using the equality 
\begin{eqnarray}
\frac{dR}{d^4Q} = \frac{dN}{d^4Xd^4Q} = \frac{dN}{\frac{1}{2}dM^2dyd^2Q_{\perp}d^2x_{\perp}\tau d\tau d y}.
\end{eqnarray}
Thus, the off-equilibrium dilepton yield can be determined using 
\begin{eqnarray}
\frac{dN}{dM^2d y} = \frac{1}{2}\int d^2Q_{\perp}d^2x_{\perp}\tau d\tau \frac{dR}{d^4Q}, 
\end{eqnarray}
where $d^2Q_{\perp} = 2 \pi Q_{\perp}dQ_{\perp}$ and the integration over $x_{\perp}$  is simply taken to be the overlapping area of the two colliding nuclei. As for the calculations of real photon production, the overlap area is estimated using Glauber Monte-Carlo results  \cite{Loizides:2017ack}. Finally,  the off-equilibrium dilepton yield can be determined from the expression 
\begin{eqnarray}
\frac{dN}{dM^2d y} &=& A_T \frac{\alpha_{EM}^2}{6\pi^4} \int \tau d\tau  \int dp_{1\perp}dp_{1z}Q_{\perp}dQ_{\perp} \\ \nonumber
&\hspace{1em}& \times \frac{p_{1\perp}}{E_1\sqrt{4Q_{\perp}^2p_{1\perp}^2-(2E E_1-M^2)^2}}  \\ \nonumber
&\hspace{1em}& \times f_q(p_{1\perp},p_{1z},\tau) \\ \nonumber
&\hspace{1em}& \times f_{\bar{q}}(\sqrt{Q_{\perp}^2+p_{1\perp}^2-2E E_1+M^2},-p_{1z},\tau ).
\end{eqnarray}

\begin{figure}[!htb]
	\begin{centering}
		\includegraphics[width=\linewidth]{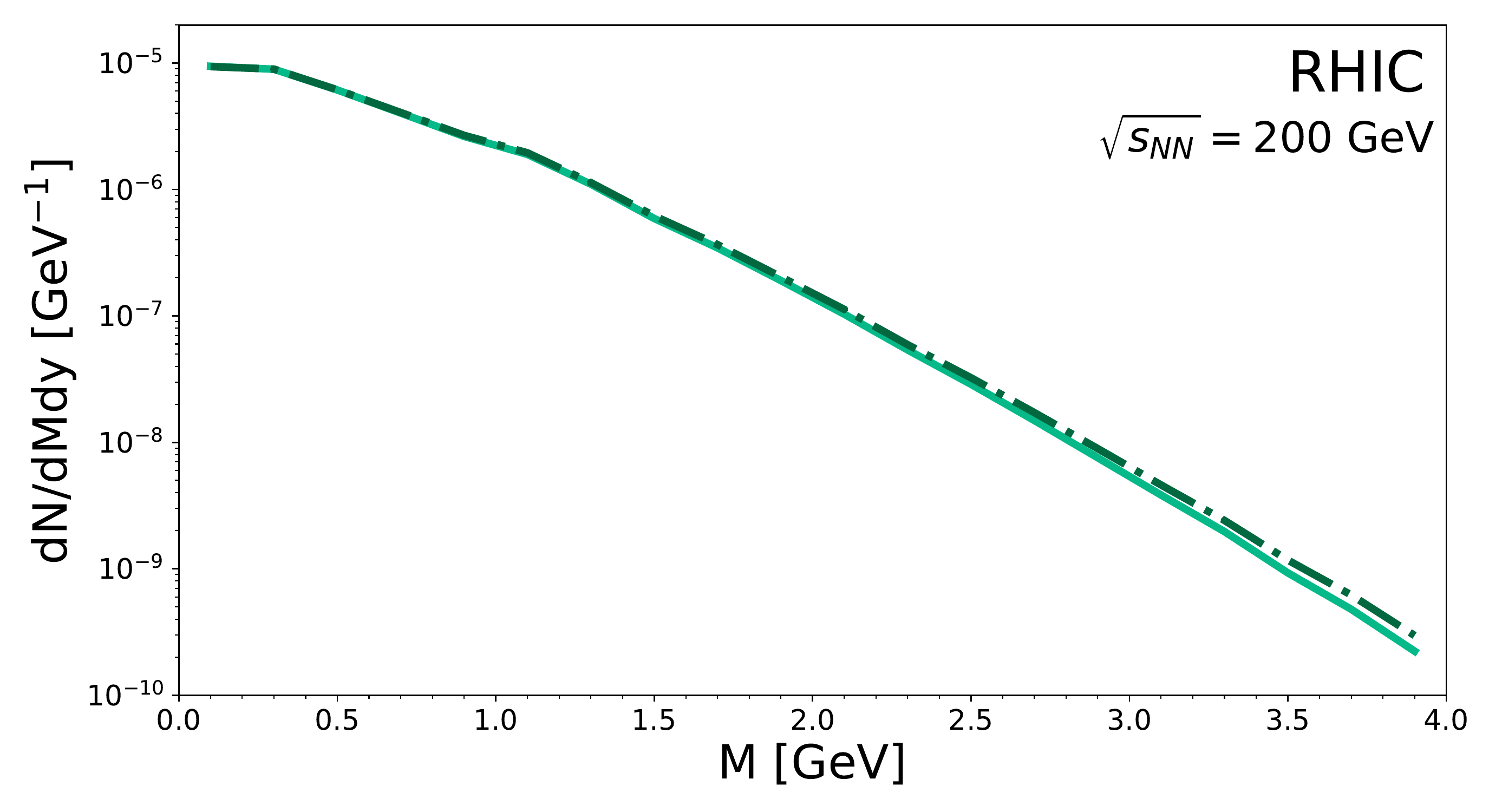}
		\includegraphics[width=\linewidth]{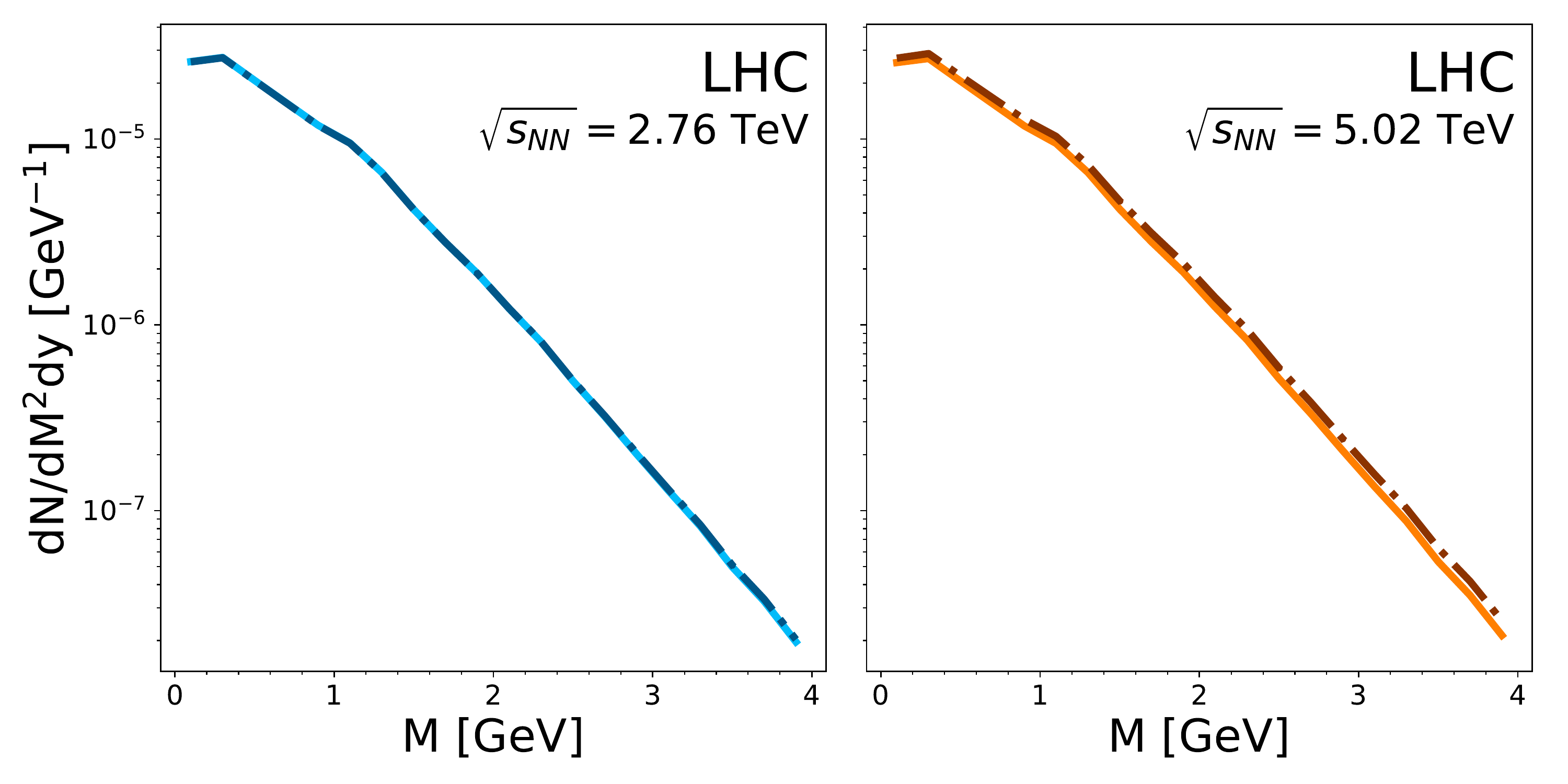}
		\caption{Pre-equilibrium dilepton yield integrated from $1/Q_s$ to 0.4 fm/c for Au+Au collisions at 200 GeV (top row). The bottom row shows the pre-equilibrium dilepton spectrum for LHC conditions: $\sqrt{s}_{NN} = 2.76$ TeV, $Q_s = 2$ GeV (left panel), and $\sqrt{s}_{NN} = 5.02$ TeV, $Q_s =2$ GeV (right panel). The solid curve is for $\xi = 1$ and the dot-dashed line is for $\xi = 1.5$.  All results reported here  are for a $0 - 20\%$ centrality class. }
		\label{fig:dilepton_RHIC_LHC}
	\end{centering}
\end{figure}
Investigating the effect of the initial gluon anisotropy on dilepton spectra, going from $\xi = 1$ to $\xi = 1.5$ we observe an increase of $\sim$10\% at RHIC ($\sqrt{s_{NN}}$ = 200 GeV, $Q_s$ = 1 GeV), and $\sim$1\% at the LHC ($\sqrt{s_{NN}}$ = 2.76 TeV and $Q_s$ = 2 GeV). For the top energy of the LHC in heavy-ion mode ($\sqrt{s_{NN}}$ = 5.02 TeV and $Q_s$ = 2 GeV) the increase due to the anisotropy is $\sim 15\%$. Those numbers are for an invariant mass $M = 2.5 $ GeV and for a centrality class $0 - 20\%$, as reported in Fig. \ref{fig:dilepton_RHIC_LHC}.  

\subsection{\ Other sources and experimental data}

In the low invariant mass region considered in this study, other contributing dilepton sources are Drell-Yan production from primordial nucleon-nucleon interactions \cite{Yan:2015zoa}.  The pairs radiated from in-medium reactions involving mesons and baryons have traditionally been considered prime sources of information on in-medium properties \cite{Rapp:1999qu,*Rapp:2009yu}. The hadrons which freeze-out at the end of the strong interaction era will also emit lepton pair via radiative decay channels. This last source is commonly referred to as ``the cocktail''. In addition, the leptons coming from semi-leptonic open-charm meson decay can combine and constitute an irreducible background \cite{Shor:1989yt} as far as the other sources are concerned. However, if the detector suite has the capability to recognize and analyze displaced decay vertices \cite{Pruneau:2017ypa}, those sources can be subtracted.

\begin{figure}[!htb]
	\begin{centering}
		\includegraphics[width=\linewidth]{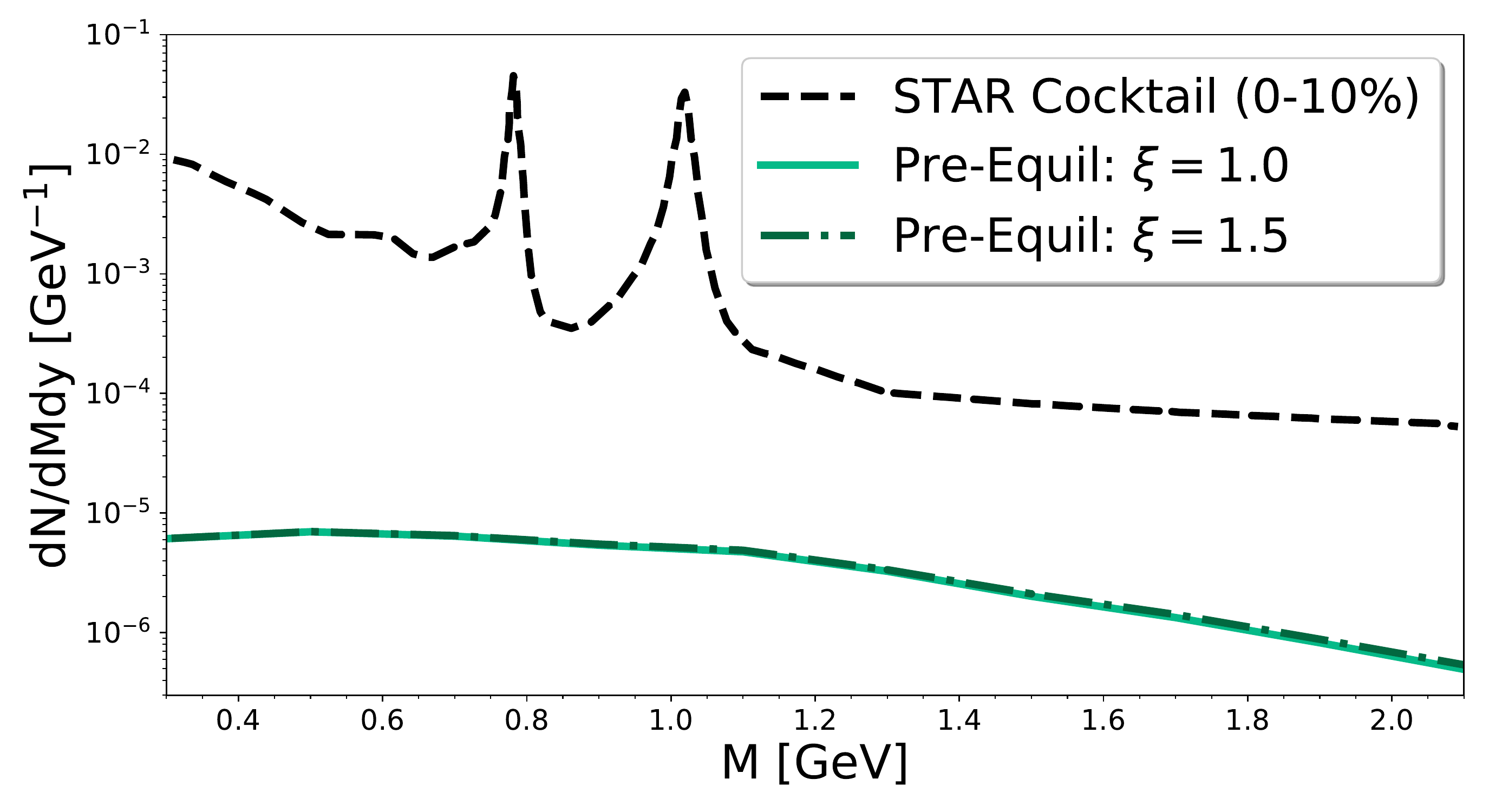}
		\caption{Pre-equilibrium dilepton yield integrated from $1/Q_s$ to 0.4 fm/c for Au+Au collisions at 200 GeV compared to the STAR Cocktail \cite{Ruan:2012za} for a 0-10\% centrality class. }
		\label{fig:dilepton_cocktail_STAR}
	\end{centering}
\end{figure}

In the case of dileptons, it turns out their revealing potential is substantially less promising, in what concerns the nature of the pre-equilibrium phase. This is situation is due to the fact that the Born graph describing lepton pair production, Figure \ref{fig:dilepton}, has an initial state which consists of quarks and anti-quarks only; no gluons at leading order. In a gluon-dominated initial state, the fermions are only produced in secondary processes, as described in Section \ref{sec:boltz}. To get a sense of  scale, it is useful to compare with the cocktail contribution which defines the threshold for new physics in the dilepton channel. This is shown on Fig. \ref{fig:dilepton_cocktail_STAR}.
Unfortunately, the dilepton signal corresponding to pre-equilibrium emission sits orders of magnitude below the cocktail. 


\section{Discussion and Conclusion} 
\label{sec:results}

This study has shown that, in the end, the potential for electromagnetic radiation to resolve the anisotropy in the initial gluon distribution is not large. This is illustrated by the photons results of Fig. \ref{fig:photon_xi} and by the dileptons results of Fig. \ref{fig:dilepton_RHIC_LHC}. We find that this statement holds at both RHIC and the LHC, at least for the systems studied in this work. This is a consequence of the fact that our model assumes an initial parton population which is gluon-dominated, with quarks and anti-quarks appearing dynamically as time evolves, as shown in Fig. \ref{fig:number_density}. The fermions hardly carry a visible imprint of the initial gluon spectrum asymmetry: this is especially visible in the dilepton channel. In addition the global pressure anisotropy is also rapidly quenched, as reported in Fig. \ref{fig:P_L/P_T}.

Our estimate of the non-equilibrium photon production has shown that the net yield can contain a measurable signature of the early time dynamics.  This is to be contrasted with the findings for lepton pairs, where the pre-equilibrium contribution is suppressed. As mentioned previously, dileptons suffer from the fact that the duration of the pre-equilibrium phase is not sufficiently long to build up an appreciable population of $q \bar{q}$ pairs. The photons will also suffer from this  paucity of fermions, but this more than compensated by the relatively large (in comparison) number of gluons. See Fig. \ref{fig:prequ_Compton_qq}. This model calculation  therefore suggests that the early, elusive,  gluon saturation distribution can inform the final photon spectrum. The influence of a gluon-dominated early hydrodynamic medium on the photon spectrum has been studied in Ref. \cite{Vovchenko:2016ijt}. 

One obvious missing element in this work is the contribution of the pre-equilibrium electromagnetic radiation to the so called ``direct photon flow puzzle'' \cite{Paquet:2015lta,David:2019wpt}. Naively, an extra early-time component should reduce the overall elliptic flow, as the net $v_2$ is a weighted average. It is the elliptic flow of each source weighted by the associated photon multiplicity. However, simply ascribing a vanishing elliptic flow to non-equilibrium photons may well neglect the possibly complicated - or even chaotic - dynamics  which can influence photons and, to a much lesser extent, hadrons.  For instance, the early-time behavior of the colored plasma is still not well understood, and consequently its modeling is the subject of much debate. The effect  on electromagnetic observables of plasma instabilities triggered by the chromoelectric field is still unclear, but recent theoretical developments carrying promising results leave hope for an answer in the near future  \cite{Hauksson:2018jog,*Hauksson:2020etn}.  Early magnetic fields can also influence electromagnetic spectra, depending on their duration and on their magnitude \cite{Basar:2012bp,Fukushima:2012fg,Tuchin:2012mf,Ayala:2019jey}. In that context, the simple dynamics used in this work do not allow for initial coordinate-space inhomogeneities in the transverse plane. For one-body observables like single particle spectra, this may be acceptable but becomes more questionable for empirical variables that depend on geometry like $v_2$, or geometry and fluctuations, like $v_3$. These elements all point out the need for a quantitative and realistic space-time picture of very early relativistic heavy-ion collisions like, for example, 3D IP-Glasma \cite{Ipp:2017uxo,*McDonald:2020oyf},  K{\o}MP{\o}ST \cite{Kurkela:2018wud,*Gale:2020xlg,*GaleKompost}, or even magnetohydrodynamics. 

Even if the investigation of early time photon spectra, real and virtual, is not yet as popular as that of later photon production, some studies have been devoted to that topic.
The pre-equilibrium photon contribution found in this work can be compared with results of the 3D Boltzmann equation simulation approach, BAMPS \cite{Greif:2016jeb}: an approach close in spirit to what is done here. At RHIC, the photon yields reported here are almost one order of magnitude above those of BAMPS, at low transverse momentum ($p_T \sim 1$ GeV), they cross at $p_T \sim 2.5$ GeV. However, BAMPS utilizes \textsc{pythia}, an initial state  very different from the simple CGC-like picture  used in this work. However, this difference may just be used to prove the point that the photon spectrum is indeed sensitive to details of the initial stages. Work which models the initial state with the Abelian Flux Tube model \cite{Oliva:2017pri} generates a partonic distribution with the Schwinger mechanism that evolves via $2 \to 2$ scattering comes close to what is performed here, even though some details differ. That reference also finds imprints of the initial conditions on the final photon spectrum, and the magnitude of the signal is comparable to the effects observed in this work. Similarly, calculations realized in the parton-hadron-string dynamics (PHSD) scenario conclude that the photons (and also dileptons, in that approach) do show characteristic features that can be attributed to specific initial state configurations \cite{Moreau:2015rrs}. 
It appears that the results reported here for the pre-equilibrium photon spectra are lower than those obtained in a realization of the bottom-up thermalization scenario calculated in Ref. \cite{Garcia-Montero:2019vju}, by a factor $\sim 2$ (LHC), $ \sim 3$ (RHIC). The bottom-up estimate for the photon yield  and the one performed herein cross at $p_T \sim 3$ GeV. Finally, recent preliminary estimates using K{\o}MP{\o}ST to model the pre-equilibrium stage \cite{GaleKompost} point to a somewhat smaller photon contribution than the one reported here, together with a slight increase in photon elliptic flow. That work is being completed and will appear soon. The partial conclusion which follows from our comparison with other results in the literature is that the evaluation of pre-equilibrium electromagnetic radiation is still a developing effort, and that results obtained so far do not rule out the exciting possibility of an experimental identification. 

To conclude, the time-dependent parton dynamics of early-time heavy-ion collisions were modeled with the relativistic Boltzmann equation, solved in the diffusion approximation. It was argued that the dynamics and non-equilibrium emission rates used here make a plausible case for an observable signature of a pre-equilibrium component in the net photon spectra.  The magnitude of the real photon signal makes a quantitative statement about kinetic and chemical equilibrium. Indeed, it appears that the fermion suppression observed at early times is largely compensated by the gluon-rich environment in the evaluation of QCD Compton scattering. One early goal of our investigation was to identify a possible electromagnetic signature of a BEC; however this has not been seen in the one-body observables considered here.   Work is ongoing to continue these investigations at higher order in $\alpha_s$, to include photon-hadron correlations, and to involve more elaborate simulation approaches.

\acknowledgments{This work was supported in part by the Natural Sciences and Engineering Research Council of Canada, and in part by the National Natural Science Foundation of China (NSFC) under Grant No. 11975079 and Shanghai Pujiang Program (No. 19PJ1401400).  We are grateful to J.-F. Paquet for providing results of his calculations of pQCD photons and of the photons generated during the hydrodynamic evolution.  C.G. acknowledges useful discussions with L. McLerran, and those following from an ongoing collaboration with   J.-F. Paquet,  B. Schenke, and C. Shen. All of us are happy to acknowledge discussions with the other members of the Nuclear Theory group at McGill University. }


\bibliography{references}

\end{document}